\documentstyle[amsmath,amssymb,graphicx]{article}

\def\be{\begin{eqnarray}}
\def\ee{\end{eqnarray}}
\def\nn{\nonumber}

\def\p{\partial}
\def\tr{{\rm tr}\,}

\hoffset= - 1.1in         

\def\cPsi{\Psi_R}
\def\cphi{{\cal X}_R}
\def\cpsi{{\cal Y}_R}

\begin{document}


\hfill ITEP/TH-20/09

\bigskip

\centerline{\Large{Unitary Integrals and Related
Matrix Models
}}

\bigskip

\centerline{Alexei Morozov}

\bigskip

\centerline{\it ITEP, Moscow, Russia}

\bigskip

\centerline{ABSTRACT}

\bigskip

{\footnotesize
Concise review of the basic properties of unitary matrix integrals.
They are studied with the help of the three matrix models:
the ordinary unitary model, Brezin-Gross-Witten model
and the Harish-\-Charndra-\-Itzykson-\-Zuber model.
Especial attention is paid to the tricky sides of the story,
from De Wit-t'Hooft anomaly in unitary integrals
to the problem of correlators with Itzykson-Zuber measure.
Of technical tools emphasized is the method of
character expansions.
The subject of unitary integrals remains highly
under-investigated and a lot of new results are expected
in this field when it attracts sufficient attention.
}



\section{Introduction}

Integrals over non-Hermitian matrices attracted far less
attention during entire history of
matrix model theory \cite{MAMOs}-\cite{KazMig}.
This is unjust, both because they are no less important
in applications than Hermitian models and because they
can be effectively studied by the same methods and are
naturally included into the unifying M-theory of
matrix models \cite{AMMmt}.
Still, in the recent reincarnation of the matrix model
theory \cite{DV}-\cite{MScs} non-Hermitian models
are once again under-investigated, and neither their
non-trivial phase structure \textit{a la} \cite{DV,AMMhe},
nor string-field-theory-like reformulation \textit{a la}
\cite{Eyft}, nor
combinatorial solution in
the Gaussian (single-cut) phase
\textit{a la} \cite{MScs} are explicitly analyzed --
despite all these subjects are clearly within reach
of the newly-developed approaches.

The present text is also too short to address these
important issues.
Instead it is concentrated on the \textit{complement}
of these \textit{universal} subjects:
on methods and results \textit{specific} to one
particular -- and most important of all non-Hermitian
models -- to that of Gaussian \textit{unitary} ensembles.
Importance of unitary integrals is obvious from the
Yang-Mills-theory perspective: they describe angular
-- color -- degrees of freedom, the complement of
diagonal -- colorless or hadronic -- components,
which are adequately captured by the eigenvalue Hermitian
models \cite{UFN3} and by their GKM
(Generalized Kontsevich Model \cite{UFN3,GKM,GKMdev})
"duals".
As already mentioned, this does not mean that various
kinds of eigenvalue techniques -- from orthogonal
polynomials \cite{MAMOs} to Virasoro constraints
\cite{virco} and their iterative \cite{Leq,AMMhe} or
QFT-style \cite{Eyft} solutions -- can not be
applied to unitary integrals -- they can and they
are  pretty effective, as usual \cite{Umo,BMS}.
This means that of the main interest are \textit{questions}
of another type, concerning the correlators of
\textit{non-diagonal} matrix elements rather than those
of traces or determinants.

A powerful -- and partly adequate -- technique for the
study of \textit{these} aspects of unitary integrals
is that of character expansions, and one of the main
intrigues in it is the interplay between the two
relevant groups: linear ($GL$) and symmetric ($S$).
This interplay is now attracting a new attention because
of the close link found between Yang-Mills theory
-- best represented for these purposes by lattice
Kazakov-Migdal model \cite{KazMig}-\cite{MMS} --
and Hurwitz theory of ramified coverings of Riemann surfaces
\cite{Hur}-\cite{AMMops}  -- represented by the Hurwitz-Kontsevich
partition function \cite{HKpf}-\cite{MMN1} and associated
peculiar matrix model \cite{MSwops,BEMS}.

This paper is no more than a brief introduction into
these aspects of unitary integrals. Our presentation
is mostly about the three basic unitary matrix models:

$\bullet$
original unitary model \cite{Umo}, describing
the correlators of traces of unitary matrices
by the standard matrix-model methods, thus we touch
it only briefly;

$\bullet$
BGW (Brezin-Gross-Witten \cite{BGW}) model,
describing the correlators of arbitrary matrix elements,
which was made -- perhaps, surprisingly at that time --
a part of the GKM theory in \cite{GKMKM,MMS}
and was finally incorporated into the M-theory of
matrix models \cite{AMMmt} in \cite{AMMbgw};

$\bullet$
IZ (Itzykson-Zuber \cite{HC,IZ}) model, a simplified
single-plaquette version of the Kazakov-Migdal theory
\cite{KazMig}, describing arbitrary correlators with
non-trivial but exactly-solvable weight (IZ action),
-- an old and difficult subject \cite{KazMigdev}-\cite{Sha},
which attracted some new attention recently in \cite{EyIZ}.

A lot of new -- and much more profound -- results are
expected in the study of unitary integrals in foreseeable
future, which -- unlike that of Hermitian integrals --
is in a rather early stage of development.
This makes the presentation of this paper essentially
incomplete and temporal -- it is no more than a
preliminary introduction.
Still, we tried to concentrate on facts and formulas,
which have good chances to remain in the core of the
subject when it becomes more complete and self-contained.
Instead some currently important -- but temporal --
results, i.e. those which will supposedly be significantly
improved and represented in more adequate terms,
are discussed in less detail and even unjustly ignored
because of the space limitations.

\section{Unitary integrals and BGW model
\label{uni}}

\subsection{The measure}

Unitary integrals are those over unitary $N\times N$
matrices $U$, $UU^\dagger = U^\dagger U = I$
with invariant Haar measure $[dU]$.
In variance with the case of Hermitian matrices
$H=H^\dagger$, where $dH = \wedge_{i,j} dH_{ij}$,
Haar measure $[dU]$ is non-trivial and non-linear in $U$.
If unitary matrices are expressed through Hermitian,
one can express $[dU]$ through $dH$, for example \cite{BMS},
\be
U = \frac{I+iH}{I-iH} \Longrightarrow
||\delta U||^2 =
4\tr\left(\frac{I}{I+iH}\,\delta H\frac{I}{I-iH}\,\delta H\right)
\ \ \Longrightarrow \ \
[dU] \sim \frac{dH}{\det ^N(1+H^2)}
\ee

A slightly more complicated is the parametrization
$U = e^{iH}$.
Then  $\ \delta U = \int_0^1 e^{isH}\delta H e^{i(1-s)H}ds$,
and
\be
||\delta U||^2 = \tr \delta U^\dagger \delta U
= -\tr (U^\dagger\delta U U^\dagger \delta U)
= \int_0^1\int_0^1 ds ds'\tr e^{i(s-s')H}\delta H e^{i(s'-s)H}\delta H
\ee
At a point where $H$ is diagonal, $H = {\rm diag}(h_1,\dots,h_N)$,
this gives
\be
||\delta U||^2 =
\sum_{i,j} |\delta H_{ij}|^2
\int_0^1\int_0^1 ds ds'\tr e^{i(s-s')(h_i-h_j)}
= \sum_{i,j} \frac{4\sin^2\frac{h_i-h_j}{2}}{(h_i-h_j)^2}\,|\delta H_{ij}|^2
\ee
This metric is already diagonal in components of $\delta H$,
therefore the measure
\be
[dU] = \prod_{i<j} \frac{4\sin^2\frac{h_i-h_j}{2}}{(h_i-h_j)^2}
\prod_{i,j} dH_{ij} =
\frac{\Delta^2_{trig}(H)}{\Delta^2(H)} dH
\ee
Since for $H = V^\dagger DV$ where $V$ is unitary and
$D = {\rm diag}(h_1,\ldots,h_n)$ diagonal,
Hermitian measure is well known to contain the square
of the Van-der-Monde determinant,
$dH = [dV]\ \Delta^2(H) \prod_i dh_i$,
we get for parametrization $U = e^{iH}$:
\be
[dU] =  [dV]\ \Delta^2_{trig}(H) \prod_i dh_i
= [dV] \prod_{i<j} 4\sin^2\frac{h_i-h_j}{2} \prod_i dh_i
\ee

\bigskip

Alternative representation is through generic complex matrix
$M$ \cite{GKMKM,MMS,EyIZ}:
\be
[dU] =  \delta(MM^\dagger - I) d^2M =
d^2M \int dS \exp\Big(i\,\tr S(MM^\dagger - I)\Big)
\label{UvsS}
\ee
where $S$ is auxiliary Hermitian matrix, used as
Lagrange multiplier to impose the unitarity constraint on $M$.
Below we often absorb $i$ into $S$, what makes it
anti-Hermitian, i.e. changes the contour of $S$-integration.
More than that, sometime one can absorb a non-trivial
\textit{matrix} factor into $S$, what completely breaks its
hermiticity, still for the integral this is often no more
than an innocent change of integration contour.

In what follows we normalize the measure $[dU]$ so that
$\int [dU] = 1$, i.e. divide $[dU]$ by the volume $V_N$
of the unitary group.
This volume can be calculated, for example, by
orthogonal polynomial method, see s.3.9 in the first
paper of ref.\cite{UFN3},
and it is somewhat sophisticated:
\be
V_N = \frac{(2\pi)^{N(N+1)/2}}{\prod_{k=1}^N k!}
\ee
We also assume that the factor $V_N^{-1}$ is included into
the definition of Hermitian matrix integrals.

\newpage

\subsection{Elementary unitary correlators
in fundamental representation}

With this normalization agreement and the symmetry properties
one immediately gets:
\be
\langle U_{ij}U^\dagger_{kl} \rangle\ =
\int U_{ij}U^\dagger_{kl} [dU] = \frac{1}{N}\delta_{il}\delta_{jk}
\label{UU+}
\ee

\be
\int U_{ij}U_{i'j'}U^\dagger_{kl}U^\dagger_{k'l'} [dU] =
\frac{N^2\Big(\delta_{il }\delta_{i'l' }\delta_{jk }\delta_{j'k' }
+ \delta_{il' }\delta_{i'l }\delta_{jk' }\delta_{j'k }\Big)
- N\Big(\delta_{il'}\delta_{i'l}\delta_{jk}\delta_{j'k'}
+ \delta_{il}\delta_{i'l'}\delta_{jk'}\delta_{j'k}\Big)}
{N^2(N^2-1)} 
\label{UUU+U+}
\ee
It is clear from this example that
the correlators with more entries become
increasingly complicated and one needs a systematic
approach to evaluate them.
This approach is provided by the theory of BGW model \cite{MMS}.

Before proceeding to this subject in s.\ref{BGWM} below,
we note that these correlators are significantly simplified
if we contract indices, i.e. consider the correlators of
\textit{traces} instead of particular matrix elements.
Indeed, from (\ref{UU+}) and (\ref{UUU+U+}) we get
\be
\int \tr U \,\tr U^\dagger\, [dU] = 1, \nn \\
\int \tr U^2 \,\tr (U^\dagger)^2\, [dU] = 2 - \delta_{N1},\nn \\
\int \tr U^2\, (\tr U^\dagger)^2\, [dU] = \delta_{N1},\nn \\
\int (\tr U)^2\, (\tr U^\dagger)^2\, [dU] = 2 - \delta_{N1}
\label{trU2}
\ee
and so on (the case of $N=1$ in these formulas
requires additional consideration, see below).
Such \textit{colorless} correlators are studied by
the ordinary unitary-matrix model \cite{BMS}.

Note that unitary correlators in (\ref{UUU+U+}) possess
a non-trivial $N$-dependence, with spurious poles at some
integer values of $N$. This phenomenon is known as
De Wit-t'Hooft anomaly \cite{DWH}.
Note also that
in correlators of colorless quantities (\ref{trU2})
these poles turn into peculiar non-singular corrections
at the same values of $N$.

\subsection{From the theory of unitary model \cite{BMS}
\label{elecor}}

Partition function of the unitary matrix model is
the generating function of correlators
of colorless trace operators:
\be
Z_U(t) = \int \exp\left(\sum_{k=-\infty}^\infty
t_k\tr U^k\right) [DU]
\ee
It was also considered in \cite{Umo}.
Note that $U^\dagger = U^{-1}$ and conjugate $U$-matrices
are represented here by negative powers of $U$.
Virasoro constraints have the form \cite{BMS}
\be
\left\{\sum_{k=-\infty}^\infty kt_k\left(
\frac{\partial}{\partial t_{k+n}}-
\frac{\partial}{\partial t_{k-n}}\right)
+ \sum_{1\leq k\leq n}\left(
\frac{\partial^2}{\partial t_k \partial t_{n-k}} +
\frac{\partial^2}{\partial t_{-k}\partial t_{k-n}}\right)\right\}
Z_U(t) = 0,
\ \ \ \ n\geq 1,
\nn \\
\left\{\sum_{k=-\infty}^\infty kt_k\left(
\frac{\partial}{\partial t_{k+n}}+
\frac{\partial}{\partial t_{k-n}}\right) + \sum_{1\leq k\leq n}\left(
\frac{\partial^2}{\partial t_k \partial t_{n-k}} -
\frac{\partial^2}{\partial t_{-k}\partial t_{k-n}}\right)\right\}
Z_U(t) = 0,
\ \ \ \ n\geq 0\
\ee
The use of these Virasoro relations is a little unusual:
there is no shift of any time variable and formal series
solution is directly in terms of $t$'s, not of their ratio
with some background values of time-variables, which
label the choice of phase, say, in Hermitian model \cite{AMM1sf}.
The analogue of the $L_0$ constraint states that
\be
\sum_{k=-\infty}^\infty kt_k\frac{\partial Z_U}{\partial t_k} =0
\ee
and requests that monomial items in $Z_U$ should have
the form $\prod_{k} t_k^{n_k}$ with
$\sum_{k\geq 1} kn_k = \sum_{k\geq 1} kn_{-k}$,
i.e. allowed are $t_kt_{-k}$, $t_kt_lt_{-k-l}$ and so on,
but not $t_k$ or $t_kt_{-k-l}$ with $l\neq 0$.
With this selection rule, the first few terms of $Z_U$ series are
\be
Z_U = e^{Nt_0} \Big(1 + a_{1|1}t_1t_{-1} + a_{2|2}t_2t_{-2} +
a_{11|2}(t_1^2 t_{-2} + t_{-1}^2t_2)
+ a_{11|11} t_1^2t_{-1}^2 + \ldots\Big)
\label{ZUexpan}
\ee
They describe the correlators (\ref{trU2}),
and are dictated by the lowest Virasoro constraints
with $n=\pm 1,\pm 2$, but in somewhat non-trivial manner:
$$
0 = e^{-Nt_0}
\left(\sum_{k=-\infty}^\infty kt_k\frac{\partial}{\partial t_{k+1}}
+ \frac{\p^2}{\p t_0\p t_1}\right)Z_U =
$$ \vspace{-0.2cm}
\be
= t_{-1}\Big(-N + N\alpha_{1|1}\Big)
+ t_1t_{-2}\Big(-2\alpha_{1|1}+\alpha_{2|2}+2N\alpha_{2|11}\Big)
+ t_1t_{-1}^2\Big(-N\alpha_{1|1}+\alpha_{2|11}+2N\alpha_{11|11}\Big)
+ \ldots
\ee
Similarly,
$$
0 = e^{-Nt_0}
\left(\sum_{k=-\infty}^\infty kt_k\frac{\partial}{\partial t_{k+2}}
+ \frac{\p^2}{\p t_0\p t_2} + \frac{\p^2}{\p t_1^2}\right)Z_U =
$$ \vspace{-0.2cm}
\be
= t_{-2}\Big(-2N + N\alpha_{2|2}+2\alpha_{2|11}\Big)
+ t_{-1}^2\Big(-\alpha_{1|1}+N\alpha_{2|11}+2\alpha_{11|11}\Big)
+ \ldots
\ee
Virasoro constraints require that each bracket at the r.h.s.
vanishes, and this implies that
\be
&&\alpha_{1|1}=1, \nn \\
(N^2-1)\alpha_{2|2} = 2(N^2-\alpha_{1|1}) = 2(N^2-1)
& \Rightarrow & \alpha_{2|2} = 2 - s\delta_{N,1}, \nn \\
(N^2-1)\alpha_{2|11} = 0 & \Rightarrow & \alpha_{2|11}
= s\delta_{N,1}\nn \\
2(N^2-1)\alpha_{11|11} = (N^2-1)\alpha_{1|1} = N^2-1
& \Rightarrow & \alpha_{11|11} = \frac{1}{4}(2-s\delta_{N,1})\nn\\
&\ldots &
\ee
Thus Virasoro constraints do not define coefficients
$\alpha_{2|2}$, $\alpha_{2|11}$ and $\alpha_{11|11}$
unambiguously -- at $N=1$.
In fact there is a single ambiguous parameter $s$
in these three coefficients, and it can be fixed by
direct evaluation of the correlators (\ref{trU2}) at $N=1$.
Similarly, coefficients $a$ with higher gradation level
$l = \sum_{k\geq 0} kn_k > 2$
are not fully defined by Virasoro constraints
at $N=1,\ldots,l$.

\subsection{From the theory of BGW model \cite{MMS}
\label{BGWM}}

Partition function of the BGW matrix model \cite{BGW} is
the generating function of all unitary correlators:
\be
Z_{BGW} = \int e^{\tr (J^\dagger U + JU^\dagger)} [dU]
= \int\int e^{iS(MM^\dagger -I)}
e^{\tr J^\dagger M + JM^\dagger} d^2M dS =\nn\\
\stackrel{(\ref{UvsS})}{=}\
\int \frac{dS}{\det S^N}\ e^{\tr (JJ^\dagger /S) - \tr S}
\ \ \ \stackrel{S\rightarrow 1/S}{=}\ \ \
\int \frac{dS}{\det S^N}\ e^{\tr (JJ^\dagger S) - \tr 1/S}
\label{ZBGW}
\ee
Obviously, $Z_{BGW}$ depends only on the eigenvalues
of the matrix $JJ^\dagger$, i.e. on the time-variables
$t_k = \frac{1}{k}\tr(JJ^\dagger)^k$.

The model has two essentially different phases, named
Kontsevich and character phases in \cite{MMS}, here
we consider only the latter one, when $Z_{BGW}$ is
a formal series in \textit{positive} powers of $t_k$.
Virasoro constraints in this phase can be derived
in the usual way: as Ward identities \textit{a la} \cite{virco},
but in this particular case they can be alternatively
considered as direct corollary of the identity
$UU^\dagger = I$ for unitary matrices, expressed in
terms of $t$-variables. This identity implies that
\be
0 = \left(\frac{\partial}{\partial \tilde J^\dagger}
\frac{\partial}{\partial J} - I\right)Z_{BGW}
= \sum_{n=1}^\infty (JJ^\dagger)^{n-1} \hat L_n Z_{BGW}
- I Z_{BGW}
\ee
i.e.
\be
\hat L_n Z_{BGW} = \left(
\sum_{k=1}^\infty kt_k\frac{\p}{\p t_{k+n}}
+ \sum_{\stackrel{a+b=n}{a,b\geq 0}}\frac{\p^2}{\p t_a\p t_b}
\right) Z_{BGW}(t) = \delta_{n,1}, \ \ \ \ n\geq 1
\ee
are the usual \textit{discrete-Virasoro} constraints
of \cite{GMMMO}, familiar from the theory of Hermitian model,
only with $n\geq 1$ instead of $n\geq -1$ and with a
modified r.h.s. at $n=1$.

Substituting a formal-series anzatz for $Z_{BGW}(t)$
into these Virasoro constraints one can iteratively
reconstruct any particular coefficient,
see s.3.2.1 of \cite{MMS}:
\be
Z_N^+(t) = 1 + \sum_{M\geq 1}\left(
\sum_{k_1\geq\ldots\geq k_M\>0} c_N\{k_a\}
\frac{k_1 t_{k_1}\ldots k_M t_{k_M}}
{(k_1+\ldots +k_M)!}\right)
\ee
and
\be
c_N\{k_a\} = \hat c_N\{k_a\} \prod_{l=0}^{k_1+\ldots+k_M-1}
\frac{1}{N^2-l^2}
\label{cNpoles}
\ee
\be
\hat c_N(1) = N, \nn \\ \nn \\
\hat c_N(2) = -N, \ \ \ \ \
\hat c_N(1,1) = N^2, \nn \\ \nn \\
\hat c_N(3) = 4N,\ \ \ \ \
\hat c_N(2,1) = -3N^2, \ \ \ \ \
\hat c_N(1,1,1) = N(N^2-2), \nn \\ \nn \\
\hat c_N(4) = -30N,\ \ \ \ \
\hat c_N(3,1) = 8(2N^2-1),\ \ \ \ \
\hat c_N(2,2) = 3(N^2+ 6),\nn \\
\hat c_N(2,1,1) = -6N(N^2-4),\ \ \ \ \
\hat c_N(1,1,1,1) = N^4-8N^2+6,\nn \\
\ldots
\ee
The first two lines of this list reproduce
eqs.(\ref{UU+}) and (\ref{UUU+U+}).

\subsection{De Wit-t'Hooft anomaly \cite{DWH,MMS}}

Remarkable feature of (\ref{cNpoles}) is the occurrence
of poles at integer values of the matrix size $N$.
It deserves noting that this is the property of the
character phase: nothing like this happens
in Kontsevich phase of BGW model -- as usual,
if properly defined, Kontsevich partition function is
actually independent of $N$ \cite{GKM,MMS}.
As we saw in s.\ref{elecor},
these poles are associated with \textit{ambiguity}
of solution to Virasoro equations in the ordinary
unitary matrix model.

In fact, DWH poles do not show up in correlators:
numerators vanish at the same time as denominators,
and singularity can be resolved by l'Hopitale rule.
However, some care is needed.
For example, in (\ref{UUU+U+}) one can consider
$$\int U_{11}^2 (U_{11}^\dagger)^2 [dU] =
\frac{2(N^2-N)}{N^2(N^2-1)} = \frac{2}{N(N+1)}
\ \stackrel{N=1}{\longrightarrow}\ 1$$
in accordance with (\ref{trU2}).
However, if one first take traces in (\ref{UUU+U+}),
and then put $N=1$, then one does not obtain
(\ref{trU2}):
$$\int \tr  U^2 \tr (U^\dagger)^2\, [dU] \
\stackrel{(\ref{UUU+U+})}{=} \
\frac{2(N^2\cdot N^2-N\cdot N)}{N^2(N^2-1)} = 2
\neq 2 - \delta_{N1}$$
Anomaly of this kind is familiar from experience with
dimensional regularizations, where taking
traces and putting dimension equal to particular value
also are non-commuting operations.


\subsection{BGW model in the M-theory of matrix models
\cite{AMMbgw}}

As explained in \cite{AMMmt} (see also \cite{Givdeco}
and \cite{AMM1sf} for preliminary results),
partition function of
every matrix model in every phase can be decomposed
into elementary constituents, belonging to Kontsevich
family of \cite{GKM}.
The most important building block is the original cubic
Kontsevich tau-function $\tau_K$, while BGW tau-function,
made from $Z_{BGW}$ in Kontsevich phase, appears to be the
next-important one
(for example, it arises, along with $\tau_K$,
in decomposition of the complex matrix model.
Both $\tau_K$ and $\tau_{BGW}$ satisfy the same
\textit{continuous-Virasoro} constraints of
\cite{virco}, 
only with $n\geq -1$ in the case of $\tau_K$ and
with $n\geq 0$ in the case of $\tau_{BGW}$.
Another, compensating, difference is that instead of
the second time-variable in the case of $\tau_K$,
for $\tau_{BGW}$ \textit{shifted} is the first time:
this allows the smaller set of constraints to unambiguously
defined the partition function.
See \cite{AMMbgw} for many more details of this
important construction.

\subsection{Other correlators and representations}

A slight modification of BGW model is to make the power of
$\det S$ in (\ref{ZBGW}) an additional free variable.
In this way we get
the celebrated Leutwyler-Smilga integral \cite{LeSm},
\be
Z_{BGW}(\nu;t) =
\int (\det U)^\nu e^{\tr (J^\dagger U + JU^\dagger)} [dU]
\label{lesm}
\ee
which plays important role in applications
and is widely discussed in the literature.


Another modification involves the change of representation:
in (\ref{ZBGW}) partition function is defined as generating
the integrals over unitary matrices in the fundamental
representation of dimension $N$ of $GL(N)$,
but one can  instead consider an arbitrary representation
$R$ of dimension $D_R(N)$.
All correlators will be, of course, related to those
in the fundamental representation, but in a somewhat non-trivial
way, and they will be different.
For example, instead of
(\ref{UU+}) in generic representation we have
\be
\int {\cal U}_{ab}{\cal U}^\dagger_{cd}\ [dU] =
\frac{\delta_{ad}\,\delta_{bc}}{D_R}
\label{UUR}
\ee
Moreover, different representations are orthogonal
in the sense that
\be
\int {\cal U}_R {\cal U}^\dagger_{R'}\ [dU]
\sim \delta_{RR'}
\label{URUR}
\ee
An important way to handle this kind of generalization is
through the comprehensive theory of character expansions.
The origins of this theory are in the fundamental works of
G.Frobenius and it is widely represented in modern theory,
from pure mathematics to applied physics.
As to references, the most we can do is to
cite just a few basic textbooks \cite{Frotext}
and \textit{just some} related papers
in the matrix-model literature \cite{charmamo,Kazcha}.
Our actual presentation will mostly follow \cite{AMMops},
while applications
of character theory to BGW and IZ models in s.\ref{IZsec}
will be discussed along
the lines of \cite{GKMKM,MMS} and \cite{Ba}.

\newpage

\section{Basics of character calculus}

\subsection{Three definitions of characters}

Character depends on representation $R$ and on conjugation class
of a group element,
which can be represented through time-variables
$t_1,t_2,\ldots$ or through auxiliary
matrix (Miwa) variable $X$, accordingly we use two notations
$\chi_R(t) = \chi_R[X]$.

Representations $R$ of $GL(\infty)$ are labeled by Young diagrams,
i.e. by ordered integer partitions
$R: \ \lambda_1\geq \lambda_2 \geq \ldots \geq 0$,
of $|R|$ -- the number of boxes in the diagram,
$\sum_j \lambda_j = |R|$, all $\lambda_j$ are integers.

Time-variables form  an infinite chain, which
-- if one likes -- can be considered as
eigenvalues of the infinite matrix from $GL(\infty)$.
Miwa variables define an $N$-dimensional subspace
in the space of time-variables,
parameterized by the eigenvalues $x_i$ of an $N\times N$ matrix $X$,
in the following way:
$t_k = \frac{1}{k}\tr X^k = \frac{1}{k}\sum_i x_i^k$.
Here and below $\tr$ without additional indices denotes
a trace of an $N\times N$ dimensional matrix, i.e. a trace in the
first fundamental representation of $GL(N)$, associated with a single-box
Young diagram $R = 1 = \Box$.  The same matrix $X$ can be converted into
arbitrary representation $R$ of $SL(N)$,
then we denote it by ${\cal X}^{(R)}$.
While $X$ can be actually considered as independent of $N$, i.e. defined
as an element of $GL(\infty)$, with $N$ just the number of non-vanishing
eigenvalues, its conversion into ${\cal X}^{(R)}$ actually depends on $N$,
thus one should be a little more accurate with formulas, which involve
${\cal X}^{(R)}$. However, for many purposes the $N$-dependence enters
only through dimensions $D_R$ of representations $R$ of $SL(N)$,
see eq.(\ref{DRdR}) below.
In other words, while ${\cal X}^{(R)}$ and $D_R$ essentially depend on $N$,
if formulas are written in terms of these objects, $N$
often does not show up.
Accordingly we suppress label $N$ in  ${\cal X}^{(R)}$ and $D_R$.

\bigskip

There are three important definitions of characters:

$\bullet$ the first Weyl determinant formula
\be
\chi_R(t) = \det_{ij}\Big(s_{\lambda_i-i+j}(t)\Big)
\label{Weyl1}
\ee
expressing $\chi_R(t)$ through determinant of Shur polynomials,
which are defined by
\be
\exp \left(\sum_k t_k z^k\right) = \sum_k s_k(t)z^k
\ee

$\bullet$ the second Weyl determinant formula
\be
\chi_R(t) = \chi_R[X] = \frac{\det_{ij} x_i^{\lambda_j+N-j}}
{\Delta(X)} = \frac{\det_{ij} x_i^{\lambda_j-j}}{\det_{ij} x_i^{-j}}
\label{Weyl2}
\ee
where Van-der-Monde determinant $\Delta(X) =
\prod_{i<j}(x_i-x_j) = \det_{ij} x_i^{N-j} = \det_{ij} x_i^{j-1}$,

$\bullet$ the trace formula
\be
\chi_R(t) = \chi_R[X] = {\rm Tr}_R \ {\cal X}^{(R)}
\label{chaXR}
\ee

\bigskip

Actually. like a Janus, these characters have two faces.
They are actually characters of two very different universal groups:
$GL(\infty)$ and $S(\infty)$, where $S_N$ is a group of permutations
of $N$ elements. Intimate relation between these two groups
is already reflected in the definition of Miwa variables,
expressing time-variables as symmetric functions of $X$-eigenvalues.

Important role in the character calculus is played by two
additional functions: dimension of representations of $SL(N)$
and $S_N$, associated with the Young diagram $R$:
\be
D_R = {\rm dim}_R\Big(GL(N)\Big)
= \chi_R\left(t_k = \frac{N}{k}\right)
= \chi_R[I],
\ \ \ \ N = |R|
\ee
and
\be
d_R  = \frac{1}{N!}\,{\rm dim}_R\Big( S(N)\Big) =
\chi_R\Big(t_k = \delta_{k1}\Big), \ \ \ \ N = |R|
\ee
Parameter $d_R$ is given by the hook formula:
\be
d_R = \prod_{\stackrel{{\rm over\ all\ boxes\ of}}
{{\rm Young\ diagram}\  R}}
\frac{1}{{\rm hook\ length}}
= \det_{1\leq i,j\leq |R|} \frac{1}{(\lambda_j+i-j)!}
\label{hoofo}
\ee
and the ratio
\be
\frac{D_R}{d_R} = \prod_{i=1} \frac{(\lambda_i + N -i)!}{(N-i)!}
\label{DRdR}
\ee
Note that $d_R$, obtained by division of dimension
by $N!=|R|!$, depends only on parameters $\lambda_i$,
while $D_R$ contains an additional explicit dependence on $N$.

One more important fact is that determinant of a matrix,
i.e. a product of all its eigenvalues, is also equal to
a character in the totally antisymmetric representation
$R = [\underbrace{1,1,\ldots,1}_N]$,
\be
\det X = 
\chi_{11\ldots 1}[X], \ \ \ \
{\rm rank}(X) = N
\label{detchar}
\ee
Moreover, integer powers of determinant are also characters:
\be
(\det X)^\nu = 
\chi_{\nu\nu\ldots \nu}[X], \ \ \ \
{\rm rank}(X) = N
\label{detcharnu}
\ee
Note that when ${\rm rank}(X) = N$ the time-variables
$t_k$ with $k>N$ are algebraic functions of those with $k\leq N$,
and, when restricted to this subspace,
the characters with $|R|>N$ also become algebraic functions of
characters with $|R|\leq N$.

\subsection{Sum rules and orthogonality relations}

Characters satisfy a number of "sum rules":
certain sums over \textit{all} representations
(i.e. over all Young diagrams or over all ordered partitions)
are equal to some distinguished quantity.
So far the most important in applications were
the following three sum rules:
\be
\sum_R d_R\chi_R(t) = e^{t_1} = e^{\tr X}
\label{1sum}
\ee
\be
\sum_R \chi_R(t)\chi_R(t') =
\exp\left(\sum_{k=1}^\infty kt_kt'_k\right)
= \frac{1}{{\rm Det}\, (I\otimes I - X\otimes X')}
\label{2sum}
\ee
\be
\sum_R \frac{d_R}{D_R}\chi_R[X]\chi_R[X'] =
\int [dU] \exp (\tr XUX'U^\dagger)
\label{sumIZ}
\ee
Eq.(\ref{1sum}) is a particular case of (\ref{2sum}),
because $d_R = \chi_R(\delta_{k1})$.
Eq.(\ref{sumIZ}) is actually a character expansion
of Itzykson-Zuber integral, but it can be effectively
used as a sum rule as well.


A very important role in the formalism of character theory
is played by orthogonality relation:
\be
\chi_R(\tilde \partial)\chi_{R'}(t) = \delta_{R,R'}
\ \ \ \ {\rm for}\ \ \ |R|=|R'|
\label{orthor}
\ee
Here one substitutes time-derivatives
$\tilde\p_k \equiv \frac{\p}{\p p_k} = \frac{1}{k}\frac{\p}{\p t_k}$
instead of $t_k$ in the arguments of $\chi_R(t)$ -- a polynomial
of its variables.
For example, see s.\ref{charexa},
\be
\chi_1(t) = t_1, & \Longrightarrow &
\chi_1(\tilde\p) = \frac{\p}{\p t_1}, \nn \\
\chi_2(t) = t_2 + \frac{t_1^2}{2}, & \Longrightarrow &
\chi_{2}(\tilde \p) = \frac{1}{2}\left(\frac{\p}{\p t_2} +
\frac{\p^2}{\p t_1^2}\right), \nn \\
\chi_{11}(t) = -t_2 + \frac{t_1^2}{2}, & \Longrightarrow &
\chi_{11}(\tilde \p) = \frac{1}{2}\left(-\frac{\p}{\p t_2} +
\frac{\p^2}{\p t_1^2}\right)
\ee
and one can easily check that (\ref{orthor}) works in these
simple cases of $|R|=1,2$. It is also clear that orthogonality
relation fails for $|R|\neq |R'|$.


\subsection{Examples of characters \label{charexa}}

For illustrative purposes we list here the first few characters
and some other relevant characteristics of the lowest representations
$R$ of $GL$ and $S$ groups.

\bigskip

\centerline{
$
\begin{array}{c|c|c|c|c|c|c|c}
&&&&&&&\\
&&d_R = &{\rm dim}_R(S_{|R|})&{\rm dim}_R(SL(n))
& \varphi_R(2) =& \varphi_R(3) = & \varphi_R(22) =\\
R & \chi_R  & \chi_R(\delta_{k,1}) &  = |R|!\,d_R
&  =\chi_R\left(\frac{n}{k}\right)
& \frac{{\rm coeff}\ t_2t_1^{|R|-2}}{2d_R}
& \frac{{\rm coeff}\ t_3t_1^{|R|-3}}{3d_R}
& \frac{{\rm coeff}\ t_2^2t_1^{|R|-4}}{2^2d_R} \\
&&&&&&&\\
\hline
&&&&&&&\\
(0) & 1 & 1 & 1 & 1 & 0 & 0 & 0 \\
&&&&&&&\\
\hline
&&&&&&&\\
(1) & t_1 & 1 & 1 & n & 0 & 0& 0 \\
&&&&&&&\\
\hline
&&&&&&&\\
(2) &
t_{2} + \dfrac{t_{1}^2 }{2}
& \frac{1}{2} & 1 & \frac{n(n+1)}{2} & 1 & 0 & 0\\
&&&&&&&\\
(1,1) & -t_{2}+\dfrac{t_{1}^2 }{2}
& \frac{1}{2} & 1 & \frac{n(n-1)}{2} &  - 1 & 0 & 0\\
&&&&&&&\\
\hline
&&&&&&&\\
(3)& t_{3}+t_{2} t_{1} + \dfrac{t_1^3}{6}
& \frac{1}{6} & 1 & \frac{n(n+1)(n+2)}{6} & 3 & 2 & 0\\
&&&&&&&\\
(2,1) &  -t_{3}+\dfrac{t_1^3}{3}
& \frac{1}{3} & 2 & \frac{n(n^2-1)}{3} & 0 & -1 & 0 \\
&&&&&&&\\
(1,1,1) & t_{3}-t_{2} t_{1} +\dfrac{t_1^3}{6}
&\frac{1}{6} & 1 & \frac{n(n-1)(n-2)}{6} & -3 & 2 & 0 \\
&&&&&&&\\
\hline
&&&&&&&\\
(4) &
t_{4}+ t_{3} t_{1} +\dfrac{t_2^2}{2} +\dfrac{t_2t_1^2}{2} +\dfrac{t_1^{4}}{24}
& \frac{1}{24} & 1 & \frac{n(n+1)(n+2)(n+3)}{24} & 6 & 8 & 3 \\
&&&&&&&\\
(3,1) &
-t_{4}-\dfrac{t_2^2}{2} +\dfrac{t_2t_1^2}{2} +\dfrac{t_1^4}{8}
& \frac{1}{8} & 3 & \frac{(n-1)n(n+1)(n+2)}{8} & 2 & 0 & -1 \\
&&&&&&&\\
(2,2) & t_{2}^2-t_{3} t_{1}+\dfrac{t_1^4}{12}
& \frac{1}{12} & 2 & \frac{n^2(n^2-1)}{12} &  0 & -4 & 3 \\
&&&&&&&\\
(2,1,1) & t_{4}-\dfrac{t_2^2}{2} -\dfrac{t_2t_1^2}{2} +\dfrac{t_1^4}{8}
& \frac{1}{8}& 3 & \frac{(n+1)n(n-1)(n-2)}{8} & - 2 & 0 & -1\\
&&&&&&&\\
(1,1,1,1) &  -t_{4}+ t_{3} t_{1} +\dfrac{t_2^2}{2} -\dfrac{t_2t_1^2}{2} +\dfrac{t_1^{4}}{24}
& \frac{1}{24} & 1 & \frac{n(n-1)(n-2)(n-3)}{24} & -6 & 8 & 3 \\
&&&&&&&\\
\hline
&&&&&&&\\
\ldots &&&&& &&\\
&&&&&&&\\
(k) &  & \frac{1}{k!} & 1 & \frac{(n+k-1)!}{(n-1)!k!}
& \frac{k(k-1)}{2} & \frac{k(k-1)(k-2)}{3} & \frac{k(k-1)(k-2)(k-3)}{8} \\
&&&&&&&\\
\ldots &&&&&&& \\
&&&\sum_R {\rm dim}_R^2 &&&&\\
&&&= |R|!&&&& \\
\end{array}
$
}

\newpage

\subsection{Example: eq.(\ref{chaXR}) for $GL(2)$}

Construction of ${\cal X}^{(R)}$ from a given $X$ is a little tricky.
Actually,
in explicit form eq.(\ref{chaXR}) is a relation for unitary matrices:
\be
\chi_R\left[ e^{A}\right] = {\rm Tr}_R\ e^{{\cal A}^{(R)}}
\ee
and relation is simple between $A$ and ${\cal A}^{(R)}$:
the $n\times n$ and $D_R\times D_R$ matrices, both expressed
through the same set of parameters $u_a$:
$$A = i\sum_{a=0}^{n^2-1} u_a T_a^{(1)}
\ \ \ \ {\rm and} \ \ \ \ {\cal A}^{(R)} =
i\sum_{a=0}^{n^2-1} u_a T_a^{(R)}$$
Here $T_a^{(R)}$ are generators of
the $GL(n)$ algebra in representation $R$,
and $R=1=\Box$ is the first fundamental representation.

For example, in the case of $GL(2)$
\be
A= \left(\begin{array}{cc} a+d & b \\ c & a-d \end{array}\right)
\nn\ee
and we have:
\be
\chi_1 \left[ e^{A}\right] = t_1 = {\rm Tr}_{D_1=2} \ e^{A}
= {\rm Tr}_{2}\ \left\{\exp \left(\begin{array}{cc}
a+d & b  \\ c& a-d\end{array}\right)\right\}
\nn\ee
\be
\chi_{2} \left[ e^{A}\right] = t_2 + \frac{t_1^2}{2} =
\frac{1}{2}{\rm Tr}_2\, e^{2A}
+ \frac{1}{2}\left({\rm Tr}_2\, e^{A}\right)^2
= {\rm Tr}_{D_2=3}\ \left\{\exp \left(\begin{array}{ccc}
2a+2d & b & 0 \\ 2c & 2a & 2b \\ 0 & c & 2a-2d \end{array}\right)\right\}
\nn\ee
\be
\chi_{11} \left[ e^{A}\right] = -t_2 + \frac{t_1^2}{2} =
-\frac{1}{2}{\rm Tr}_2\, e^{2A}
+ \frac{1}{2}\left({\rm Tr}_2\, e^{A}\right)^2 = e^{2a}, \ \ \ D_{11}=1
\nn\ee
\be
\chi_{3} \left[ e^{A}\right] = t_3 + t_2t_1 + \frac{t_1^3}{6} =
\frac{1}{3}{\rm Tr}_2\, e^{3A}
+ \frac{1}{2}\left({\rm Tr}_2\, e^{2A}\right)\left({\rm Tr}_2\, e^{A}\right)
+ \frac{1}{6}\left({\rm Tr}_2\, e^{A}\right)^3 = \nn \\
= {\rm Tr}_{D_3=4}\ \left\{\exp \left(\begin{array}{cccc}
3a+3d & b & 0 & 0 \\ 3c & 3a+d & 2b & 0 \\ 0 & 2c & 3a-d & 3b \\
0&0&c&3a-3d\end{array}\right)\right\}
\nn\ee
\be
\chi_{21} \left[ e^{A}\right] = -t_3 + \frac{t_1^3}{3} =
- \frac{1}{3}{\rm Tr}_2\, e^{3A}
+ \frac{1}{3}\left({\rm Tr}_2\, e^{A}\right)^3 =
{\rm Tr}_{D_{21}=2}\ \left\{\exp \left(\begin{array}{cc}
3a+d & b  \\ c& 3a-d\end{array}\right)\right\}
\nn\ee
\be
\chi_{111} \left[ e^{A}\right] = t_3 - t_2t_1 + \frac{t_1^3}{6} =
\frac{1}{3}{\rm Tr}_2\, e^{3A}
- \frac{1}{2}\left({\rm Tr}_2\, e^{2A}\right)\left({\rm Tr}_2\, e^{A}\right)
+ \frac{1}{6}\left({\rm Tr}_2\, e^{A}\right)^3 = 0, \ \ \ D_{111}=0
\nn\ee
\be
\ldots
\nn\ee

Note that:

$\bullet$
matrix $A$ does not actually need to be anti-Hermitian,
above relations are actually true for arbitrary $A$;

$\bullet$
normalization of the central $U(1)$ generator includes a
representation-dependent factor $|R|$;

$\bullet$
in general, for representation of the size $r\times r$
of $GL(2)$ we have:
\be
T_+ = \left(\begin{array}{cccccc}
0 & 1 & 0 &  & 0 & 0 \\
0 & 0 & 2 & & 0 & 0 \\ &&&\ldots &&\\
0&0&0&&r-2&0\\ 0&0&0&&0&r-1\\ 0&0&0&&0&0
\end{array}\right) \ \ \ \ \ \ \
T_- = \left(\begin{array}{cccccc}
0 & 0 && 0 & 0 & 0 \\
r-1 & 0 && 0 & 0 & 0 \\
0&r-2&&0&0&0\\
&&\ldots &&&\\
0&0&&2&0&0\\
0&0&&0&1&0
\end{array}\right) \nn \\
T_3 = \left(\begin{array}{ccccc}
r-1 & 0 &&  0 & 0 \\
0 & r-3 &&  0 & 0 \\
&&\ldots &&\\
0&0&&3-r&0\\
0&0&&0 & 1-r
\end{array}\right) \ \ \ \ \ \ \
T_0 = |R|\left(\begin{array}{ccccc}
1 & 0 &&  0 & 0 \\
0 & 1 &&  0 & 0 \\
&&\ldots &&\\
0&0&&1&0\\
0&0&&0&1
\end{array}\right)
\ee
with commutation relations
$[T_+,T_-] = T_3$, $[T_3,T_\pm] = \pm 2 T_\pm$.

\subsection{Characters and symmetric functions
\cite{GKMKM,Ba}}

Many application of characters are based on the fact
that they are symmetric functions (of eigenvalues $x_i$ of a matrix $X$),
and simultaneously are eigenfunctions and eigenvalues
of certain distinguished operators.

In the space of symmetric functions there are three "obvious"
sets of homogeneous generators:
\be
{\rm time-variables} \ \ \ \ \
p_k[X] = \sum_{i} x_i^k = \tr X^k, \ \ \ \ k\geq 0, \ \ \ \ \
p_k[X] = kt_k[X],
\ee
\be
{\rm Shur\ polynomials} \ \ \ \ \
s_k[X] =  \sum_{i_1\leq\ldots\leq i_k} x_{i_1}\ldots x_{i_k},
\ \ \ \ k\geq 1, \ \ \ \ s_0[X] = 1
\ee
and
\be
{\rm elementary\ symmetric\ polynomials}\ \ \ \ \
\sigma_k[X] = \sum_{i_1<\ldots<i_k} x_{i_1}\ldots x_{i_k},
\ \ \ \ k\geq 1, \ \ \ \ \sigma_0[X] = 1.
\ee
The corresponding three generating functions are
\be
P(z) = \sum_{k=0}^\infty p_k[X] z^k = \tr \frac{I}{I-zX}
\ee
\be
S(z) = \sum_{k=0}^\infty s_k[X] z^k =
\prod_i (1-zx_i)^{-1} = \frac{1}{\det(I-zX)}
\ee
and
\be
\Sigma(z) = \sum_{k=0}^\infty \sigma_k[X] z^k =
\prod_i (1+zx_i) = \det(I+zX)
\ee
These three generating functions are related as follows:
\be
S(z) = \exp \left(\sum_{k=1}^\infty t_k[X]z^k\right) =
\frac{1}{\det(I-zX)}
\ee
where $p_k[X] = kt_k[X]$, so that
\be
P(z) = \tr I + z\frac{\partial}{\partial z} \log E(z)
\ee
and
\be
\Sigma(z) = 1/S(-z)
\ee

The sets $\{t_k\}$, $\{s_k\}$ and $\{\sigma_k\}$ are generators
in the space of symmetric functions: any such function is
a \textit{poly}-linear combination of generators.
\textit{Linear} bases are formed by products of these generators
and are labeled by Young diagrams $R$.
An obvious choice of basis is
\be
t^R \equiv \prod_j t_{\lambda_j} \equiv \prod_j t_j^{m_j}
\ \ \ \ {\rm or} \ \ \ \
p^R \equiv \prod_j p_{\lambda_j} \equiv \prod_j p_j^{m_j} =
\left(\prod_j j^{m_j}\right) t^R
= {\rm const}\cdot t^R, \ \ \ \
t_\emptyset = p_\emptyset = 1
\label{pR}
\ee
In this way one can define the "$R$-power" of any
semi-infinite sequence $\{p_i\}$, not obligatory from
the time-variables.
Sometime we need the same basis with a different
normalization:
\be
\widetilde{p^R} \equiv \frac{p^R}{z_R}
\ \ \ \
{\rm with} \ \ \ \
z_R \equiv \prod_j m\!_j\,! j^{m_j}
\label{ptR}
\ee
Note that widetilde here is \textit{not} exactly the same as tilde
in orthogonality relation (\ref{orthor}).

A less trivial linear basis is provided by characters,
made from the generators $\{s_k\}$ with the help of (\ref{Weyl1}),
\be
\chi_R = \det_{ij} \left(s_{\lambda_j-j+i}\right)
\ee
Motivation for this choice comes from the following expansion rule
\cite{Ba}:
\be
\prod_i \left(\sum_{k=0}^\infty A_k x_i^k\right)
= \sum_R  \chi_R[X] \cdot
\det_{ij} \left(A_{\lambda_j-j+i}\right)
\label{prodi}
\ee
which is related through the second Weyl formula (\ref{Weyl2})
to the basic determinant expansion \cite{GKMKM,Ba}
\be
\det_{ij} \left(\sum_k A_{ik}B_{jk}\right)
= \sum_{P\in S_N} (-)^P \prod_{i=1}^N \sum_k A_{ik}B_{P(i)k}
= \sum_{k_1,k_2,\ldots,k_N} \sum_P (-)^P
\prod_{i=1}^N A_{ik_i}B_{P(i)k_i} = \nn \\ =
\sum_{k_1>k_2>\ldots} \left(\sum_{P,P'} (-)^P(-)^{P'}
\prod_{i=1}^N A_{P'(i)k_i}B_{P(i)k_i}\right)
= \sum_{k_1>k_2>\ldots} \det_{ij} A_{ik_j} \det_{ij} B_{ik_j}
\label{ABprod}
\ee
which generalizes $\det (AB) = \det A\cdot \det B$ from square
to rectangular matrices.
Indeed, with $k_j = \lambda_j-j$
(note that from $\lambda_1\geq \lambda_2\geq \ldots$
it follows that $k_1=\lambda_1-1>k_2=\lambda_2-2>\ldots$)
and with (\ref{Weyl2}) substituted for $\chi_R[X]$,
the r.h.s. of (\ref{prodi})
has exactly the same form as the r.h.s. of (\ref{ABprod}).
Therefore
\be
\sum_R  \chi_R[X] \cdot
\det_{ij} \left(A_{\lambda_j-j+i}\right)
\ \stackrel{(\ref{Weyl2})}{=}\
\sum_{\lambda_1\geq\lambda_2\geq\ldots} \frac{
\det_{ij} \left(x_i^{\lambda_j-j}\right)
\det_{ij} \left(A_{\lambda_j-j+i}\right)}
{\det_{ij} x_j^{-i}} = \nn \\
\stackrel{(\ref{ABprod})}{=}\
\frac{\det_{ij} \left(\sum_k A_{k+i}x_j^k\right)}
{\det_{ij} x_j^{-i}} =
\frac{\det_{ij} \left(A(x_j)x_j^{-i}\right)}
{\det_{ij} x_j^{-i}} = \prod_j A(x_j)
\ee
with $A(x) = \sum_{k=0}^\infty A_k x^k$ and we deduce
(\ref{prodi}) from (\ref{ABprod}).
Note that summation limits are automatically adjusted
in appropriate way.

\subsection{Characters as eigenfunctions and eigenvalues
\cite{MSwops,MMN1}}

Characters $\chi_R[X]$ are the common eigenfunctions
of the infinite set of commuting operators $\hat W(R)$ \cite{MMN1}.
These operators are in one-to-one correspondence with
the Young diagrams (ordered integer partitions)
and are made from the $GL$ generator
$\hat D_{ij} = X_{ik}\frac{\partial}{\partial X_{jk}}$
in two steps: first define a semi-infinite sequence
of \textit{commuting} operators $\hat D_k =\ :\tr \hat D^k:\ $
and then use the general rule (\ref{ptR}) to introduce
\be
\hat W(R) = \ : \widetilde{D^R}:\ = \
: \prod_k \frac{1}{ m_k! k^{m_k}}\big(\tr\hat D^k\big)^{m_k} :
\ee
Additional ingredient is the normal ordering, which implies
that all $X$ stand to the left of all $\p/\p X$.
According to \cite{MMN1} these operators form the center
of universal enveloping algebra,
\be
\left[ \hat W(R), \hat W(R')\right] = 0
\ \ \ \ \ \forall R,R'
\ee
and characters are their eigenfunctions
\be
\hat W(R) \chi_{R'}[X] = \varphi_{R'}(R) \chi_{R'}[X]
\ \ \ \ \ \forall R,R'
\label{Wchi}
\ee
The corresponding eigenvalues $\varphi_R(\Delta)$
are in fact characters of symmetric group $S$,
and they can be defined as the coefficients of
$\chi_R$ expansion in $p$-variables,
\be
\chi_R(p) = \sum_{|\Delta|=|R|} d_R\varphi_R(\Delta) p^\Delta
\label{chivsp}
\ee
The sum here is over Young diagrams $\Delta$ of the same
size (number of boxes) as $R$. For $|\Delta|>|R|$
these eigenvalues vanish, while for
$|\Delta|<|R|$ they are defined as
\be
\varphi_R(\Delta,\underbrace{1,\ldots,1}_k) =
\frac{(|R|-|\Delta|)!}{\big(|R|-|\Delta|-k\big)!\,k!}\
\varphi_R(\Delta,\underbrace{1,\ldots,1}_{|R|-|\Delta|})
\ \ \ \ \ {\rm if}\ \ \ \ 1\notin \Delta
\label{phik1}
\ee
i.e. if $\Delta$ does not contain $1$ entries at all.
In particular, for such $\Delta$ we have
$\varphi_R(\Delta) =
\varphi_R(\Delta,\underbrace{1,\ldots,1}_{|R|-|\Delta|})$.
In other words, if $\Delta$ contains exactly $k$ unit entries,
then
\be
\varphi_R(\Delta) =
\frac{(|R|-|\Delta|+k)!}{\big(|R|-|\Delta|\big)!\,k!}\
\varphi_R(\Delta,\underbrace{1,\ldots,1}_{|R|-|\Delta|})
\ \ \ \ \ {\rm if}\ \ \ \ 1^k \in \Delta
\label{phik2}
\ee

Important in applications is also inverse of (\ref{chivsp}):
\be
\widetilde{p^\Delta} = \sum_{|R|=|\Delta|}d_R\chi_R(p)\varphi_R(\Delta)
\label{pvschi}
\ee
Note that $\widetilde{p\,^\Delta}$ appears at the l.h.s. instead
of $p^\Delta$ at the r.h.s. of (\ref{chivsp}). Note also that
the both formulas are written in terms of $p$ rather than $t$-variables,
$p_k = kt_k$.

Operators $\hat W(R)$ and thus their eigenvalues
form an interesting commutative associative algebra \cite{MMN1,AMMops}
\be
\hat W(R_1)\hat W(R_2) = \sum_{R_3} C_{R_1R_2}^{R_3} \hat W(R_3)
\ \ \ \Longrightarrow \ \ \
\varphi_R(R_1)\varphi_R(R_2)
= \sum_{R_3} C_{R_1R_2}^{R_3} \varphi_R(R_3)
\ \ \ \ \ \ \forall R
\ee


\section{The theory of IZ integral
\label{IZsec}}

The very important application of character theory
is the theory of Harish-Chandra-Itzykson-Zuber
integral \cite{HC,IZ}.
As mentioned in the Introduction, it is the simplified
version -- or the main basic block --
of the Kazakov-Migdal model \cite{KazMig}-\cite{MMS}
of lattice gluodynamics
and, actually, of many other matrix-model-based
approaches to non-perturbative dynamics of gauge
theories with propagating (particle-like) degrees
of freedom.
It is also a non-trivial example of group integral,
which satisfies the requirements of Duistermaat-Heckman
theorem \cite{DHth} and can be treated exactly by the methods,
now known as \textit{localization technique}
(with Nekrasov's instanton calculus \cite{Nekin}
in Seiberg-Witten theory \cite{SWTh} as one of
its remarkable achievements).

HCIZ integrals can be handled by a variety of methods,
starting from the Ward identities of \cite{IZ}.
We present here the character-based approach \cite{GKMKM,Ba}
and a closely related GKM-based technique of
\cite{KazMigdev}-\cite{MMS}, recently-revived in \cite{EyIZ}
and applied to the old (and still unsolved) problem of
arbitrary correlators with the HCIZ weight \cite{Mofo,Sha}.

\subsection{Itzykson-Zuber integral and duality formula}

IZ integral is a unitary integral with non-trivial weight
of the special form:
\be
J_{IZ}(X,Y) = \int e^{\tr X U Y U^\dagger} [dU]
\label{IZdef}
\ee
It depends on the eigenvalues of two $N\times N$ matrices $X$
and $Y$ through the celebrated determinant formula \cite{IZ}
\be
J_{IZ} \sim \frac{\det_{ij} e^{x_i y_j}}
{\Delta(X)\Delta(Y)} = \frac{1}{\Delta(X)\Delta(Y)}
\sum_{P\in S_N} (-)^P \exp\left(x_i y_{P(i)}\right),
\label{IZdet}
\ee
where $\Delta(X) = \prod_{i<j}(x_i-x_j) =
\det_{ij} x_i^{j-1}$
and $\Delta(Y)$ are the two Van-der-Monde determinants.

The standard way to derive (\ref{IZdet}) is to observe
that $J_{IZ}$ depends only on the eigenvalues of $X$ and $Y$
and substitute a function of eigenvalues into
the obvious Ward identities:
\be
\left\{
\tr \!\!\left(\frac{\partial}{\partial X^{tr}}\right)^k - \tr Y^k
\right\} J_{IZ}(X,Y) = 0, \ \ \ \ \ \
\left\{
\tr \!\!\left(\frac{\partial}{\partial Y^{tr}}\right)^k - \tr X^k
\right\} J_{IZ}(X,Y) = 0, \ \ \ \ \ \ k\geq 0
\ee
what reduces the problem to Calogero-Dunkl-like equations,
since, for example,
\be
\tr \!\!\left(\frac{\partial}{\partial X^{tr}}\right)^2
= \sum_i \frac{\p^2}{\p x_i^2} + \sum_{i\neq j}
\frac{1}{x_i-x_j}\left(\frac{\p}{\p x_i} - \frac{\p}{\p x_j}\right)
\ee
when it acts on a function of eigenvalues.
One can then check that (\ref{IZdet}) is indeed a solution.
This method is rather tedious and the answer is in no way obvious,
at least for people unfamiliar with Calogero-Dunkl equations
and associated theory of "zonal spherical functions" \cite{zsf}.
Instead it has non-trivial generalizations, a recent one \cite{MoShID}
is to the theory of \textit{integral discriminants}:
the homogeneous non-Gaussian integrals -- an important branch
of \textit{non-linear algebra} \cite{nolal}.

Direct way from (\ref{IZdef}) to the answer (\ref{IZdet})
is provided by Duistermaat-Heckman approach \cite{DHth},
which claims that an integral invariant under an action
of a compact group and with appropriate relation between the
classical action and quantum measure -- conditions, which
are trivially satisfied by the group-theoretical integral
(\ref{IZdef}), -- is exactly given by its \textit{full}
quasiclassical approximation: a sum over \textit{all}
extrema of the classical action:
\be
\int d\phi e^{-S(\phi)} \sim \sum_{\phi_0:\ \delta S(\phi_0)=0}
\left(\det\frac{\p^2S(\phi_0)}{\p\phi^2}\right)^{-1/2}
e^{-S(\phi_0)}
\label{DH}
\ee
As explained in s.3.2 of \cite{Banff}, see also \cite{EyIZ},
the equations of motion for the action
$\tr XUYU^\dagger$,
\be
\left[ X, UYU^\dagger \right] = 0,
\ee
after a unitary transformation which makes $X$ and $Y$ diagonal
have any permutation matrix, $U=P$, as their solution
(more solutions arise when some eigenvalues of $X$ or $Y$ coincide,
but this does not affect the answer).
Then Duistermaat-Heckman theorem (\ref{DH}) implies that
\be
J_{IZ}(X,Y) \sim \sum_P
\frac{(-)^P}{\Delta(X)\Delta(Y)}
\exp\left(\sum_i x_iy_{P(i)}\right)
\label{DHIZ}
\ee
where Van-der-Monde determinants arise from the pre-exponential
factor and the sign factor $(-)^P$ naturally arises from
the square root and distinguishes the contributions of
minima and maxima to the quasiclassical sum.
Dusitermaat-Heckman method, known also as \textit{localization}
technique, got a variety of applications since it was first
discovered in the study of HCIZ integrals.
A new promising application should be to the \textit{3-algebras}
and associated BLG theory \cite{BLG}, where this is still the only
available direct approach to \textit{defining} a substitute
of matrix models -- since no direct counterpart of \textit{matrices}
is yet known for 3-algebras.


Of certain importance is also the following representation
of (\ref{IZdef}) as a Hermitian-matrix integral
\cite{KazMigdev,MMS,EyIZ}.
Immediately from the definition (\ref{IZdef}) we get:
\be
J_{IZ} \ \stackrel{(\ref{IZdef})}{=}\
\int e^{\tr X U Y U^\dagger} [dU]
\ \stackrel{(\ref{UvsS})}{=} \
\int d^2M dS
e^{\tr S(I-MM^\dagger) + \tr XMYM^\dagger}
= \int \frac{e^{\tr S} dS }{\det_{N^2\times N^2}
(S\otimes I - X\otimes Y)}
\label{JIZS}
\ee
One can easily convert to the basis where, say, $Y$
is diagonal, $Y = {\rm diag}\{y_k\}$. Then
\be
J_{IZ}(X,Y) \sim \int \frac{e^{\tr S} dS}{\prod_k \det (S-y_k X)}
\ \stackrel{\cite{EyIZ}}{\longrightarrow}\
\int \frac{e^{\tr SX} dS}{\prod_k \det(S-y_kI)}
\sim
\int J_{IZ}(X,S)\, \frac{\prod_{i\neq j} (s_i-s_j)\prod_i ds_i}
{\prod_{i,k}(s_i-y_k)}
\label{JIZSev}
\ee
where determinants and traces are now over $N\times N$ matrices
and arrow transition results from the change of variables
$S \rightarrow SX$
(which implies a change of integration contour over $S$).
The last step is transition to eigenvalues of $S$,
with angular integration given again by the same
Itzykson-Zuber integral.
According to \cite{EyIZ} this duality relation
provides an effective recursion
in the size $N$ of the unitary matrix.

\subsection{Character expansion of the HCIZ integral}


From character calculus we can immediately derive an
alternative formula, called character expansion:
making use of (\ref{1sum}) one can express $e^{\tr \Psi}$ with
$\Psi \equiv X U Y U^\dagger$ as a linear
combination of characters,
$e^{\tr \Psi} \ \stackrel{(\ref{1sum})}{=}\ \sum_R d_R\chi_R(\Psi)$
and after that
substitute the average of character over the unitary group:
\be
\int \chi_R[\Psi]\ [dU] \ \stackrel{(\ref{chaXR})}{=}\
\int {\rm Tr}_R {\cPsi}\, [dU]
\ \stackrel{(\ref{chaXR})}{=}\
\int{\rm Tr}_R\Big({\cphi}{\cal U}{\cpsi}{\cal U}^\dagger\Big)[dU]
\ \stackrel{(\ref{UUR})}{=}\
\frac{1}{D_R}\chi_R[X]\chi_R[Y]
\label{charint}
\ee
Thus
\be
J_{IZ} = \int e^{\tr \Psi} [dU]
\ \stackrel{(\ref{1sum})}{=}\
\sum_R d_R \int  \chi_R[\Psi]\ [dU]
\ \stackrel{(\ref{charint})}{=}\
\sum_R \frac{d_R\chi_R[X]\chi_R[Y]}{D_R}
\label{IZchar}
\ee
This elementary but powerful formula can be found, for example,
in \cite{Ba} and -- with erroneously omitted $d_R$-factor --
in \cite{Kazcha}.
For its recent application to exact evaluation
of correlators in Gaussian Hermitian model see the last papers
in ref.\cite{MShHZ}.
Similarly one obtains a character expansion in a more
sophisticated case, with two terms in the exponent,
known as (generalized) Berezin-Karpelevich integral \cite{BeIZ,Ba}:
\be
J_{BK}=\int [dU] \int [dV] \exp
\left( X U Y V^\dagger + VY U^\dagger X \right)
= \sum_{R,R'}
d_Rd_{R'} \int\int \chi_R(X UY V^\dagger)
\chi_{R'}(VY U^\dagger X) [dU][dV] = \nn \\
\stackrel{(\ref{URUR})}{=}\ \sum_{R}  d_R^2 \int\int
{\rm Tr}_R\Big({\cphi}{\cal U}{\cpsi}{\cal V}^\dagger\Big)
{\rm Tr}_{R}\Big({\cphi}{\cal V}{\cpsi}{\cal U}^\dagger\Big)[dU][dV]
\ \stackrel{(\ref{UUR})}{=}\
\sum_R \frac{d_R^2}{D_R^2} \chi_R[X^2]\chi_R[Y^2]
\ \ \ \ \
\label{IZIZchar}
\ee
because, as a consequence of (\ref{URUR}) we have,
in addition to
\be
\int \chi_R[U]\chi_{R'}[U]\ [dU] = \delta_{R,R'}
\label{orthochar}
\ee
For applications to two-matrix models see, for example,
\cite{Kazcha}.

Representation (\ref{IZdet}) follows from (\ref{IZchar})
if one expresses characters at the r.h.s. through the
eigenvalues with the help of the second Weyl formula (\ref{Weyl2}):
\be
J_{IZ}\ \stackrel{(\ref{IZchar})}{=}\
\sum_R \frac{d_R\chi_R[X]\chi_R[Y]}{D_R}
\ \stackrel{(\ref{Weyl2})}{=}\
\frac{1}{\Delta(X)\Delta(Y)}\sum_R\frac{d_R}{D_R}
\det_{ij} x_i^{N-j+\lambda_j} \det_{ij} y_i^{N-j+\lambda_j}
= \nn \\ \stackrel{(\ref{DRdR})}{=} \
\frac{1}{\Delta(X)\Delta(Y)}
\sum_{\lambda_1\geq \lambda_2 \geq \ldots \geq 0}\ \
\prod_{i=1}^N \frac{(N-i)!}{(\lambda_i + N -i)!}
\det_{ij} x_i^{N-j+\lambda_j} \det_{ij} y_i^{N-j+\lambda_j}=\nn\\
\ \ \ \stackrel{\lambda_i + N -i = k_i}{=}\ \ \
\frac{\prod_{i=1}^N (N-i)!}{\Delta(X)\Delta(Y)}
\sum_{k_1> k_2 > \ldots }
\frac{\det_{ij} x_i^{k_j} \det_{ij} y_i^{k_j}}{\prod_j k_j!}
= \left(\prod_{j=0}^{N-1} j!\right) \frac{\det_{ij} e^{x_iy_j}}
{\Delta(X)\Delta(Y)}
\label{JIZchaIZ}
\ee
since, as a particular case of (\ref{ABprod}),
\be
\sum_{k_1> k_2 > \ldots }
\left(\prod_{j=1}^N a_j\right)
\det_{ij} x_i^{k_j} \det_{ij} y_i^{k_j}
= \frac{1}{N!}\sum_{P,P'}
\sum_{k_j}\Big(\phi_{P(j)}\psi_{P'(j)}\Big)^{k_j}
= \det_{ij} A(\phi_i\psi_j)
\label{aprod}
\ee
for $A(x) = \sum_k a_kx^k$, see \cite{Ba}.

Similarly from (\ref{IZIZchar}) we obtain another celebrated
formula \cite{BeIZ,Ba}:
\be
J_{BK}\!\! \stackrel{(\ref{IZIZchar})}{=}
\sum_R \frac{d_R^2\chi_R[X^2]\chi_R[Y^2]}{D_R^2}
= \frac{\left(\prod_{j=0}^{N-1} j!\right)^2}
{\Delta(X^2)\Delta(Y^2)}
\sum_{k_1> k_2 > \ldots }\!\!\!\!
\frac{\det_{ij} x_i^{2k_j} \det_{ij} y_i^{2k_j}}
{\left(\prod_j k_j!\right)^2}
\stackrel{(\ref{aprod})}{=}
\left(\prod_{j=0}^{N-1} j!\right)^2\!\!\!
\frac{\det_{ij} {\cal I}_0(2x_iy_j)}
{\Delta(X^2)\Delta(Y^2)}
\ee
where Bessel function ${\cal I}_0(2x) = \sum_k \frac{x^{2k}}{(k!)^2}$.
This is a kind of a transcendental generalization of the IZ
integral.

\subsection{Reductions to BGW model and Leutwyler-Smilga integral
\label{lesmchar}}

Similar character expansions exist, of course, in the
simpler case of BGW model.
However, since integral (\ref{ZBGW}) contains two items
in the exponent, it is in fact closer to the transcendental
Berezin-Karpelevich integral, than to the IZ integral.
Instead of (\ref{IZIZchar}) we get for (\ref{ZBGW}):
\be
Z_{BGW} =
\int  e^{\tr (J^\dagger U + JU^\dagger)} [dU]
= \sum_{R,R'} d_Rd_{R'} \int
\chi_R[J^\dagger U]\chi_{R'}[JU^\dagger]\ [dU]\
\stackrel{(\ref{URUR})}{=}
\sum_{R} \frac{d_R^2}{D_R}\chi_R[JJ^\dagger]
\ee
To make the next step \cite{Ba} we now need not only
(\ref{DRdR}) but also the hook formula (\ref{hoofo}):
\be
Z_{BGW} = \sum_{R} \frac{d_R^2}{D_R}\chi_R[JJ^\dagger] =
\frac{\prod_{j=0}^{N-1} j!}
{\Delta[JJ^\dagger]}
\sum_{k_1> k_2 > \ldots }
\frac{\det_{ij} J_i^{2k_j}}
{\left(\prod_j k_j!\right)}\det_{ij}\frac{1}{(k_j-N+i)!}
= \nn \\
\stackrel{(\ref{ABprod})}{=}
\left(\prod_{j=0}^{N-1} j!\right)
\frac{\det_{ij} \Big(J_j^{N-i}{\cal I}_{i-N}(2J_j)\Big)}
{\Delta(JJ^\dagger)}
= \left(\prod_{j=0}^{N-1} j!\right)
\frac{\det_{ij} \Big(J_i^{j-1}{\cal I}_{j-1}(2J_i)\Big)}
{\Delta(JJ^\dagger)}
\ee
where $J_i^2$ are eigenvalues of Hermitian matrix $JJ^\dagger$ and
the Bessel function
\be
{\cal I}_s(2x) = \sum_{k=0}^\infty \frac{x^{2k+s}}{k!(k+s)!}
= {\cal I}_{-s}(2x)
\ee

Leutwyler-Smilga integral (\ref{lesm}) can be handled
in the same way, but one should make use of appropriate
generalization of (\ref{1sum}) and (\ref{hoofo}):
\be
(\det U)^\nu e^{\tr U} = \sum_{\lambda_1\geq\lambda_2\geq\ldots\geq 0}
\chi_{\lambda_1\ldots\lambda_N}[U]
\det_{ij} \frac{1}{(\lambda_j-\nu+i-j)!}
\ee
From this formula we get \cite{LeSm,Ba}
$$
Z_{BGW}(\nu;t) =
\int (\det U)^\nu e^{\tr (J^\dagger U + JU^\dagger)} [dU]
= \frac{\prod_{j=0}^{N-1} j!}
{\Delta[JJ^\dagger]}
\sum_{k_1> k_2 > \ldots }
\frac{\det_{ij} J_i^{2k_j}}
{\left(\prod_j k_j!\right)}\det_{ij}\frac{1}{(k_j-N-\nu+i)!} =
$$ \vspace{-0.3cm}
\be
= \left(\prod_{j=0}^{N-1} j!\right)
\frac{\det_{ij} \Big(J_i^{j-1}{\cal I}_{\nu+j-1}(2J_i)\Big)}
{\Delta(JJ^\dagger)}
\ee

\subsection{Correlators of $\Psi$-variable
in IZ integral \cite{AMMops}}

Character expansion allows one to evaluate arbitrary correlators
of traces of $\Psi  = XUYU^\dagger$ and its powers in IZ integral.

For $\Delta = \Big\{\delta_1\geq\delta_2\geq\ldots\Big\} =
\Big\{ \underbrace{m,\ldots,m}_{\alpha_m},\ \ldots\ ,
\underbrace{1,\ldots,1}_{\alpha_1}\Big\}$ of the size
$|\Delta| = \sum_k \delta_k = \sum_m m\alpha_m$ we have from
(\ref{pvschi}):
%
%
\be
\Psi^\Delta \equiv
(\tr \Psi)^{\alpha_1}(\tr \Psi^2)^{\alpha_2}\ldots(\tr \Psi^m)^{\alpha_m}
\ \stackrel{(\ref{pvschi})}{=}\
z_\Delta \sum_{|R|=|\Delta|} d_R \chi_R(\Psi)\varphi_R(\Delta)
\label{psidechar}
\ee
Instead of (\ref{IZchar}) we can now write
\be
\int \Psi^\Delta e^{\tr \Psi} [dU] =
\sum_{n=0}^\infty \frac{1}{n!}
\Psi^{\Delta + n\cdot 1} [dU]
\ \stackrel{(\ref{psidechar})}{=}\
\sum_n \frac{z_{\Delta+n\cdot 1}}{n!}
\sum_{|R|=|\Delta|+n} d_R \varphi_R(\Delta + n\cdot 1)
\int \chi_R(\Psi)[dU] = \nn \\
\stackrel{(\ref{charint})}{=} \
\sum_R \frac{z_{\Delta+(|R|-|\Delta|)\cdot 1}}{(|R|-|\Delta|)!}
\frac{d_R}{D_R}
\,\varphi_R\Big(\Delta + (|R|-|\Delta|)\cdot 1\Big)\chi_R[X]\chi_R[Y]
\label{psicor}
\ee
Note that sum over $n$ is actually traded for the
sum over \textit{all} $R$, then only the contribution
with $n = |R|-|\Delta|$ is picked up from the sum over $n$.
If $\Delta$ does not contain any unit entries, then
$z_{\Delta+(|R|-|\Delta|)\cdot 1} = (|R|-|\Delta|)!\,z_\Delta$
and this sophisticated formula simplifies a little:
\be
\int \widetilde{\Psi^\Delta} e^{\tr \Psi} [dU] =
\sum_R \frac{d_R}{D_R}
\,\varphi_R\Big(\Delta + (|R|-|\Delta|)\cdot 1\Big)\chi_R[X]\chi_R[Y]
\ \ \ \ \ {\rm if} \ \ \ \ 1\notin \Delta
\ee

As a byproduct of this calculation we obtain a possible proof
of (\ref{Wchi}) \cite{AMMops}.
For this purpose we act with the operator $\hat W(\Delta) =
\widetilde D^\Delta$ on $J_{IZ}$.
Since operator $\hat D_{ij} = \sum_k X_{ik}\frac{\p}{\p X^{jk}}$
converts $\Psi_{kl} = (XUYU^\dagger)_{kl}$
into itself, $\hat D_{ij} \Psi_{kl} = \delta_{jk}\Psi_{il}$,
we have:
\be
\hat W(\Delta) J_{IZ}(X,Y) =
\hat W(\Delta) \int e^{\tr \Psi}[dU] =
\int \widetilde{\Psi^\Delta} e^{\tr \Psi} [dU]
\ee
and now it remains to substitute
(\ref{charint}) for $J_{IZ}$ at the l.h.s. and
(\ref{psicor}) for the average at the r.h.s.
Picking up the coefficient in front of $\frac{d_R\chi_R[Y]}{D_R}$
in sums over $R$ at both sides we obtain:
\be
\hat W(\Delta)\chi_R(X) =
\frac{z_{\Delta+(|R|-|\Delta|)\cdot 1}}{(|R|-|\Delta|)!\,z_\Delta }
\,\varphi_R\Big(\Delta + (|R|-|\Delta|)\cdot 1\Big)\chi_R[X]
\ \stackrel{(\ref{phik2})}{=}\ \varphi_R(\Delta)\chi_R[X]
\ee
The last equality is trivially true when $\Delta$ does not contain
unit entries, $1 \notin \Delta$, since in this case
$z_{\Delta+(|R|-|\Delta|)\cdot 1}=(|R|-|\Delta|)!\,z_\Delta$
and $\varphi_R\Big(\Delta + (|R|-|\Delta|)\cdot 1\Big) =
\varphi_R(\Delta)$.
If $\Delta$ contains $k$ unit entries, then
$z_{\Delta+(|R|-|\Delta|)\cdot 1} = \frac{(|R|-|\Delta|+k)!}{k!}z_\Delta$,
while $\varphi_R\Big(\Delta + (|R|-|\Delta|)\cdot 1\Big)
\ \stackrel{(\ref{phik2})}{=}\
\frac{k!\,(|R|-|\Delta|)!}{(|R|-|\Delta|+k)!}\,\varphi_R(\Delta)$.

Coming back to (\ref{psicor}), it provides a character expansion
for generic correlator of $\Psi$-traces.
Of course, this is only a half-stuff formula:
an analogue of the transformation (\ref{JIZchaIZ})
is still needed to convert it into a more explicit expression.
Unfortunately, in any case the correlators of $\Psi$-traces are
not the most general ones needed in HCIZ theory, and evaluation of
other correlators remains an unsolved, underestimated
and under-investigated problem to which we devote a
separate short section.

\subsection{Pair correlator in HCIZ theory}

Evaluation of unitary integrals with HCIZ measure
became an important scientific problem since the
appearance of Kazakov-Migdal model \cite{KazMig}
of lattice gluodynamics and was originally addressed
in \cite{KazMigdev} and \cite{Mofo,Sha}.
Given the Duistermaat-Heckman nature of HCIZ integral
(\ref{DHIZ}) \cite{Banff,EyIZ},
this problem is also a natural chapter in development
of "localization" techniques and BRST-like approaches
\cite{DHth}-\cite{zsf}.
Of course, the problem is also related to generic
studies of non-Gaussian integrals in the theory of
integral discriminants \cite{MoShID},
which is an important part of the newly emergent field
of \textit{non-linear algebra} \cite{nolal}.
Unfortunately, the subject of HCIZ correlators
is still underestimated
and remains poorly investigated -- not a big surprise,
given the lack of knowledge about the ordinary
unitary integrals, overviewed in s.\ref{uni} above.
Still, the interest is returning to this field,
in particular the relatively recent papers \cite{EyIZ}
provide a new confirmation and serious extension
of the old result of \cite{Mofo}.
This section provides nothing more than a sketchy review
of the present state of affairs.

Correlators can be in principle evaluated directly --
by choosing one's favorite parametrization of unitary
matrices and explicitly evaluating all the $N^2$ integrals.
A clever and convenient parametrization \cite{Sha}
makes use of Gelfand-Zeitlin variables, which are
unconstrained (free),  it also arises
naturally in recursive procedure in $N$ of \cite{EyIZ},
which starts from duality relation (\ref{JIZSev}).
Though seemingly natural, this approach does not
provide any general formulas, at least in any straightforward
way. As usual, a more practical approach is the "stringy" one
\cite{UFN2} -- that of the generating functions.

The generating function of interest in HCIZ theory is
\be
Z_{IZ} = \int e^{\tr X U Y U^\dagger
+ \tr J^\dagger U + \tr JU^\dagger} [dU]
\ \stackrel{(\ref{UvsS})}{\sim} \nn \\ \sim
\int d^2M dS \exp\Big(\tr S(I-MM^\dagger)\Big)
\exp\Big(\tr XMYM^\dagger + \tr J^\dagger M + JM^\dagger\Big)
= \nn \\
= \int \frac{e^{\tr S} dS }{\det_{N^2\times N^2}
(S\otimes I - X\otimes Y)}
\exp\left( \tr_{N^2\times N^2} J^\dagger
\Big(S\otimes I - X\otimes Y\Big)^{-1} J\right)
= \nn \\ = \int \frac{dS}{\prod_k \det (S-y_k X)}
\exp \left\{\sum_k \left( S_{kk} + J^\dagger_{ki}
(S-Xy_k)^{-1}_{ij} J_{jk} \right)\right\}
\label{ZIZ}
\ee
-- as direct generalization of (\ref{JIZS})
and (\ref{JIZSev}).
Integral over $M$ is Gaussian and Wick theorem is applicable,
non-trivial part is evaluation of the $S$ integral.
In can be interesting to include also the Leutwyler-Smilga
factor $(\det U)^\nu$ into the integrand, but
-- as we saw in s.\ref{lesmchar} --
this leads to non-trivial modifications.
Unfortunately not much is known about this function even at
$\nu =0$. There are no serious obstacles for going beyond the
first non-trivial correlator
$\langle U_{ij}U_{kl}^\dagger \rangle_{IZ}$,
still it remains the only well-studied example.

This pair correlator simplifies considerably if we
consider it in the basis where $X$ and $Y$ are both
\textit{diagonal} -- then \cite{KazMigdev}
\be
\int U_{ij}U_{kl}^\dagger \ [dU] =
{\cal M}_{ij} \delta_{kj}\delta_{il}
\ee
According to \cite{Mofo}, \textit{in this basis}
the generating function
\be
\sum_{i,j=1}^N a_i{\cal M}_{ij}\, b_j =
\sum_{i,j=1}^N a_i b_j \int |U_{ij}|^2
\exp\left(\sum_{k,l} x_ky_l|U_{kl}|^2\right) [dU]
= \frac{1}{\Delta(X)\Delta(Y)}\sum_{P\in S_N} (-)^P
e^{x_iy_{P(i)}}\cdot
\label{Mofo}
\ee
$$
\cdot\sum_{n=0}^{N-1}(-)^n
\sum_{1\leq i_1<\ldots<i_{n+1}\leq N}
\frac{\det \left(\begin{array}{ccc}
1 &\ldots & 1\\
x_{i_1} &\ldots & x_{i_{n+1}} \\
\vdots && \vdots \\
x_{i_1}^{n-1}& \ldots& x_{i_{n+1}}^{n-1} \\
a_{i_1}& \ldots &a_{i_{n+1}} \end{array}\right)
\det \left(\begin{array}{ccc}
1 &\ldots & 1\\
y_{P(i_1)} &\ldots & y_{P(i_{n+1})} \\
\vdots && \vdots \\
y_{P(i_1)}^{n-1}& \ldots &y_{P(i_{n+1})}^{n-1} \\
b_{P(i_1)}& \ldots &b_{P(i_{n+1})} \end{array}\right)}
{\det \left(\begin{array}{ccc}
1 &\ldots & 1\\
x_{i_1} &\ldots & x_{i_{n+1}} \\
\vdots && \vdots \\
x_{i_1}^{n} &\ldots& x_{i_{n+1}}^{n} \end{array}\right)
\det \left(\begin{array}{ccc}
1 &\ldots & 1\\
y_{P(i_1)} &\ldots & y_{P(i_{n+1})} \\
\vdots && \vdots \\
y_{P(i_1)}^{n} &\ldots &y_{P(i_{n+1})}^{n} \end{array}\right)}=
$$
$$
= \frac{1}{\Delta(X)\Delta(Y)}\sum_{P\in S_N} (-)^P e^{x_iy_{P(i)}}
\left( \sum_{i=1}^N a_ib_{P(i)}\ -\ \sum_{1\leq i<j\leq N}
\frac{(a_i-a_j)(b_{P(i)}-b_{P(j)})}{(x_i-x_j)(y_{P(i)}-y_{P(j)})}
+ \phantom{\frac{\left|\begin{array}{ccc}
1&1&1\\ x_i&x_j&x_k \\ a_i &a_j&a_k \end{array}\right|
}{(x_i-x_j)(x_j-x_k)(x_i-x_k)}}
\right.
$$ $$
\left. \ \ \ \ \ \ \ \ \ \
+ \sum_{1\leq i<j<k\leq N}\frac{\left|\begin{array}{ccc}
1&1&1\\ x_i&x_j&x_k \\ a_i &a_j&a_k \end{array}\right|
\left|\begin{array}{ccc}
1&1&1\\ y_{P(i)}&y_{P(j)}&y_{P(k)} \\ b_{P(i)} &b_{P(j)}&b_{P(k)}
\end{array}\right|}{(x_i-x_j)(x_j-x_k)(x_i-x_k)(y_{P(i)}-y_{P(j)})
(y_{P(j)}-y_{P(k)})(y_{P(i)}-y_{P(k)})}\ \ \  + \ \ldots
 \right)
$$
This sophisticated formula was transformed in
\cite{EyIZ} into a more elegant form:
\be
{\cal M}_{ij} =
\frac{1}{\Delta(X)\Delta(Y)}\
{\rm res}\left.
\left\{\det E - \det\left(E -
\frac{1}{X-u\cdot I}\,E\,\frac{1}{Y-v\cdot I}\right)
\right\}\right|_{\stackrel{u=x_i}{v=y_j}}
\label{Mofores}
\ee
where matrix $E$ is defined as $E_{ij} = e^{x_iy_j}$.
According to this residue formula
one should pick up the coefficients in front of
$\Big((u-x_i)(v-y_j)\Big)^{-1}$ in the sum
\be
-\frac{1}{\Delta(X)\Delta(Y)}
\sum_{P\in S_N} (-)^P \prod_k e^{x_ky_{P(k)}}
\left(1 - \frac{1}{(u-x_k)(v-y_{P(k)})}\right)
\ee
over permutations.
The first term $\det E$ in (\ref{Mofores}) does not contribute
to the residue.
There will be two different kinds
of distributions: one for permutations $P$ with the
property $j=P(i)$ and another -- for all other
permutations:
\be
{\cal M}_{ij} =
\frac{1}{\Delta(X)\Delta(Y)}
\sum_{P\in S_N} \sum_{P} (-)^P
\left\{\delta_{P(i),j}
\prod_{k\neq i}^N \left(1 - \frac{1}{(x_i-x_k)(y_j-y_{P(k)})}\right)
+ \right. \nn \\
\left. - \Big(1-\delta_{P(i),j}\Big)
\frac{1}{(x_i-x_{P(j)})(y_j-y_{P(i)})}\prod_{k\neq i,\,P^{-1}(j)}^N
\left(1 - \frac{1}{(x_i-x_k)(y_j-y_{P(k)})}\right)
\right\}\prod_{k=1}^N e^{x_k y_{P(k)}}
\ee

In particular, for $N=1$ we have a single term of the first kind
and ${\cal M}_{11} = e^{x_1y_1}$.

For $N=2$ the terms of both types contribute:
\be
{\cal M}_{11} = \frac{1}{x_{12}y_{12}}
\left\{ \left(1-\frac{1}{x_{12}y_{12}}\right)e^{x_1y_1+x_2y_2}
+ \frac{1}{x_{12}y_{12}}\,e^{x_1y_2+x_2y_1}
\right\} = {\cal M}_{22}, \nn \\
{\cal M}_{12} = \frac{1}{x_{12}y_{12}}\left\{
\frac{1}{x_{12}y_{12}}\,e^{x_1y_1+x_2y_2} -
\left(1+\frac{1}{x_{12}y_{12}}\right)e^{x_1y_2+x_2y_1}
\right\}
= {\cal M}_{21}
\label{MN2}
\ee
what is exactly the same as (\ref{Mofo}),
where just the first two terms survive at the r.h.s. for $N=2$.
Of course, ${\cal M}_{11}+{\cal M}_{12} = J_{IZ}$.

In general, equivalence of (\ref{Mofo}) and (\ref{Mofores})
follows from the fact that for $a_i = 1/(u-x_i)$
the determinants in the numerator in (\ref{Mofo}) turns into
\be
\det \left(\begin{array}{ccc}
1 &\ldots & 1\\
x_{i_1} &\ldots & x_{i_{n+1}} \\
\vdots && \vdots \\
x_{i_1}^{n-1}& \ldots& x_{i_{n+1}}^{n-1} \\
(u-x_{i_1})^{-1}& \ldots & (u-x_{i_{n+1}})^{-1} \end{array}\right)
 = \det \left(\begin{array}{ccc}
1 &\ldots & 1\\
x_{i_1} &\ldots & x_{i_{n+1}} \\
\vdots && \vdots \\
x_{i_1}^{n} &\ldots& x_{i_{n+1}}^{n} \end{array}\right)
\prod_{k=1}^{n+1} \frac{1}{u-x_{i_k}}
\ee
-- simply because the numerator should have dimension
$n(n-1)/2$ and be antisymmetric in all $x_{i_k}$.
Therefore (\ref{Mofo}) implies that
$$
\sum_{i,j} \frac{1}{u-x_i}{\cal M}_{ij} \frac{1}{v-y_j}
= \frac{1}{\Delta(X)\Delta(Y)}\sum_{P\in S_N} (-)^P
e^{x_iy_{P(i)}}\left\{\sum_{n=0}^{N-1}(-)^n
\!\!\!\!\! \sum_{1\leq i_1<\ldots<i_{n+1}\leq N}\
\prod_{k=1}^{n+1} \frac{1}{(u-x_{i_k})(v-y_{P(i_k)})}\right\}=
$$
\be
= \frac{1}{\Delta(X)\Delta(Y)}\sum_{P\in S_N}\!\! (-)^P
e^{x_iy_{P(i)}}\! \left\{1 - \prod_{i=1}^N
\left(1 - \frac{1}{(u-x_{i})(v-y_{P(i)})}\right)\right\}
= \frac{\det E -
\det\left(E - \frac{1}{u-X}E\frac{1}{v-Y}\right)
}{\Delta(X)\Delta(Y)}
\label{Mofores2}
\ee

Note that taking a limit to (\ref{UU+}) in this formula is
a somewhat tricky exercise -- this is not a big surprise,
since even $J_{IZ}(X,Y=0) = 1$ is not a fully obvious
corollary of eq.(\ref{IZdet}).
Much simpler is to check that (\ref{Mofores2}) is consistent
with $\sum_j |U_{ij}|^2 = 1$, i.e. that
$\sum_j {\cal M}_{ij} = J_{IZ}$ for any $i$.
Indeed,
\be
\sum_{i,j} \frac{1}{u-x_i}{\cal M}_{ij}
= \sum_{j} \oint_{y_j} dv
\frac{\det E -
\det\left(E - \frac{1}{u-X}E\frac{1}{v-Y}\right)
}{\Delta(X)\Delta(Y)} =
\ee
$$
= \frac{\det E}{\Delta(X)\Delta(Y)}
\oint_\infty dv \left(1 -
\det \left(1 - \frac{1}{u-X}E\frac{1}{v-Y}E^{-1}\right)\right)
= J_{IZ} \oint_\infty
\tr \frac{1}{u-X}E\frac{dv}{v-Y}E^{-1} = J_{IZ}\ \tr\frac{1}{u-X}
$$

A character decomposition of pair correlator can be easily
obtained similarly to eq.(\ref{charint}),
provided one knows appropriate generalization of
(\ref{UUR}) and (\ref{URUR}), namely
\be
\int {\cal U}_R{\cal U}^\dagger_R
{\cal U}_{R'}{\cal U}^\dagger_{R'} [dU]
\ee
-- a generalization of (\ref{UUU+U+}) to arbitrary
representations. Actually, the case when $R'$ is the
fundamental representation would be enough for decomposition
of (\ref{Mofo}).

Like HCIZ integral itself, the pair correlator satisfies
a kind of duality relation -- a direct generalization of
(\ref{JIZSev}): from
\be
\int U_{ij} U^\dagger_{kl} [dU] \ \stackrel{(\ref{ZIZ})}{\sim}\
\delta_{jk} \int \left(\frac{1}{S-Xy_k}\right)^{-1}_{il}
\frac{e^{\tr S}dS}{\prod_k\det (S-Xy_k)}
\ee
it follows, after the change of integration matrix-variable
$S \longrightarrow XS$ and after diagonalization of $X$ and $S$
that
\be
{\cal M}_{ij}(X,Y) \sim
\int\frac{ {\cal M}_{ij}(X,S)\ \Delta^2(S)\, \prod_p ds_p}
{(s_i-y_j)\,x_j\ \prod_{p,q}(s_p-y_q)}
\label{Mdu}
\ee
Similar relations can be straightforwardly deduced from (\ref{ZIZ})
for higher correlators in HCIZ theory.

For generalizations to other simple Lie groups see,
for example, \cite{BreH} and \cite{EyIZ}.

\subsection{$U(2)$ example}

To illustrate the use of duality relations we demonstrate
explicitly how (\ref{Mdu}) is satisfied by (\ref{Mofo}) and
(\ref{Mofores}) in the simplest non-trivial case of $N=2$.

We begin with an even simpler story: duality relation
(\ref{JIZSev}) at $N=1$. The statement is simply that
\be
e^{xy} = \oint \frac{e^{xs}ds}{s-y}
\ee
and it is clear that integral should be understood as taken
around the pole at $s=y$.

For $N=2$ the same (\ref{JIZSev}) becomes slightly more involved:
\be
\frac{e^{x_1y_1+x_2y_2} - e^{x_1y_2+x_2y_1}}{x_{12}y_{12}}
\ \stackrel{(\ref{ZIZ})}{\sim}\
\oint\oint
\frac{e^{x_1s_1+x_2s_2} - e^{x_1s_2+x_2s_1}}{x_{12}s_{12}}
\frac{s_{12}^2 ds_1ds_2}{(s_1-y_1)(s_1-y_2)(s_2-y_1)(s_2-y_2)}
= \nn \\
= \frac{1}{x_{12}y_{12}^2} \oint\oint
\left(e^{x_1s_1+x_2s_2} - e^{x_1s_2+x_2s_1}\right)
\left(\frac{1}{s_1-y_1} - \frac{1}{s_1-y_2}\right)
\left(\frac{1}{s_2-y_1} - \frac{1}{s_2-y_2}\right)
s_{12} ds_1 ds_2 = \nn \\
\!= \frac{1}{x_{12}y_{12}^2}\Big(
-y_{12}\!\left(e^{x_1y_1+x_2y_2}\! - e^{x_1y_2+x_2y_1}\right)
\!-y_{21}\left(e^{x_1y_2+x_2y_1}\! - e^{x_1y_1+x_2y_2}\right)
\Big) =
-2\,\frac{e^{x_1y_1+x_2y_2} - e^{x_1y_2+x_2y_1}}{x_{12}y_{12}}
\label{IZduN2}
\ee
Note that from the four possible poles,
$(s_1,s_2) = (y_1,y_1),\ (y_1,y_2),\ (y_2,y_1),\ (y_2,y_2)$
the first and the last one do not contribute because of
the $s_{12}= s_1-s_2$ factor in the numerator of the
integrand.

Coming back to (\ref{Mdu}), in the case of $N=1$
it states simply that
\be
e^{xy} = \oint \frac{e^{xs} ds}{x(s-y)^2} =
\frac{1}{x}\left.\frac{\p e^{xs}}{\p s}\right|_{s=y} = e^{xy}
\ee
For $N=2$, the calculation is much longer.
We present it only for one particular (out of four)
non-vanishing correlator (\ref{MN2}),
\be
{\cal M}_{11}(X,Y) = -\frac{1}{x_{12}^2y_{12}^2}
\Big( (1-x_{12}y_{12})e^{x_1y_1+x_2y_2} - e^{x_1y_2+x_2y_1}\Big)
\ee
With the help of the same tricks as in (\ref{IZduN2})
we can write the r.h.s. of (\ref{Mdu}) as
$$
\oint\oint \frac{s_{12}^2 ds_1ds_2}{(s_1-y_1)x_1 y_{12}^2}
\left(\frac{1}{s_1-y_1} - \frac{1}{s_1-y_2}\right)\!
\left(\frac{1}{s_2-y_1} - \frac{1}{s_2-y_2}\right)
\left\{-\!\left(\frac{1}{x_{12}^2} - \frac{s_{12}}{x_{12}}\right)
e^{x_1s_1+x_2s_2}
+ \frac{1}{x_{12}^2} e^{x_1s_2+x_2s_1}\right\}
= \nn \\
$$ $$
= \frac{1}{x_1 x_{12}^2 y_{12}^2}\oint\oint ds_1ds_2
\left\{\underline{-\frac{1}{(s_1-y_1)^2}} + \frac{1}{y_{12}}
\left(\frac{1}{s_1-y_1} - \frac{1}{s_1-y_2}\right)\right\}
\left(\frac{1}{s_2-y_1} - \frac{1}{s_2-y_2}\right)\cdot
$$ $$
\cdot
\Big((1 -s_{12}x_{12})e^{x_1s_1+x_2s_2}
- e^{x_1s_2+x_2s_1}\Big)
$$
This time all the four poles can contribute and cancellations
for two unwanted ones are a little less trivial:
for $(s_1,s_2)=(y_1,y_1)$ we have
$$
\Big(\underline{-(x_1-x_{12}) + x_2} + \frac{1}{y_{12}}(1-1)\Big)
\,e^{(x_1+x_2)y_1} = 0,
$$
while for $(s_1,s_2)=(y_2,y_2)$
$$
\frac{1}{y_{12}}(1-1)\,e^{(x_1+x_2)y_2} = 0
$$
To simplify analysis of these formulas we underline the terms,
which come from the double pole at $s_1=y_1$ and produce
derivatives of the integrand over $s_1$.
Non-vanishing are contributions from the two other poles,
from $(s_1,s_2)=(y_2,y_1)$:
$$
-\frac{1}{y_{12}}\Big((1+x_{12}y_{12})\,e^{x_1y_2+x_2y_1}
- e^{x_1y_1+x_2y_2}\Big)
$$
and, finally, from $(s_1,s_2)=(y_1,y_2)$:
$$
\underline{\Big(x_1(1-x_{12}y_{12})-x_{12}\Big)e^{x_1y_1+x_2y_2}
- x_2\,e^{x_1y_2+x_2y_1}}
- \frac{1}{y_{12}}\Big((1-x_{12}y_{12})\,e^{x_1y_1+x_2y_2}
- e^{x_1y_2+x_2y_1}\Big)
$$
Putting all together we obtain for the r.h.s. of (\ref{Mdu}):
$$
\frac{1}{x_1x_{12}^2y_{12}^2}\left\{\left(
\frac{1}{y_{12}} + \underline{x_1 -x_{12} -x_1x_{12}y_{12}}
-\frac{1}{y_{12}} +x_{12} \right)\,e^{x_1y_1+x_2y_2}
+ \left( -\frac{1}{y_{12}}-x_{12} \underline{- x_2}
+ \frac{1}{y_{12}}\right)\, e^{x_1y_2+x_2y_1}\right\} =
$$
\be
=  \frac{1}{x_{12}^2y_{12}^2}\Big\{
(1-x_{12}y_{12})\, e^{x_1y_1+x_2y_2}
- e^{x_1y_2+x_2y_1}\Big\} = -{\cal M}_{11}(X,Y)
\ee

In order to demonstrate what the higher correlators
are going to look like, we list here non-vanishing
4-point correlators for $N=2$:
\be
\int U_{11}^2(U^\dagger_{11})^2\ [dU] =
\frac{1}{x_{12}^3y_{12}^3}\Big\{
(2 - 2x_{12}y_{12} + x_{12}^2y_{12}^2)\,e^{x_1y_1+x_2y_2}
- 2\,e^{x_1y_2+x_2y_1}\Big\}, \nn \\
\int U_{11}U_{12}U^\dagger_{11}U^\dagger_{21}\ [dU] =
\frac{1}{x_{12}^3y_{12}^3}\Big\{
(-2 + x_{12}y_{12})\,e^{x_1y_1+x_2y_2}
+ (2+x_{12}y_{12})\, e^{x_1y_2+x_2y_1}\Big\},\nn \\
\int U_{11}U_{22}U^\dagger_{11}U^\dagger_{22}\ [dU] =
\frac{1}{x_{12}^3y_{12}^3}\Big\{
(2 - 2x_{12}y_{12} + x_{12}^2y_{12}^2)\,e^{x_1y_1+x_2y_2}
- 2\,e^{x_1y_2+x_2y_1}\Big\}, \nn \\
\int U_{11}U_{22}U^\dagger_{12}U^\dagger_{21}\ [dU] =
\frac{1}{x_{12}^3y_{12}^3}\Big\{
(2 - x_{12}y_{12})\,e^{x_1y_1+x_2y_2}
- (2+x_{12}y_{12})\, e^{x_1y_2+x_2y_1}\Big\}, \nn \\
\int U_{12}^2(U^\dagger_{21})^2\ [dU] =
\frac{1}{x_{12}^3y_{12}^3}\Big\{
2\,e^{x_1y_1+x_2y_2}
- (2+2x_{12}y_{12}+x_{12}^2y_{12}^2)\, e^{x_1y_2+x_2y_1}\Big\},\nn\\
\int U_{12}U_{21}U^\dagger_{12}U^\dagger_{21}\ [dU] =
\frac{1}{x_{12}^3y_{12}^3}\Big\{
2 \,e^{x_1y_1+x_2y_2}
- (2+2x_{12}y_{12}+x_{12}^2y_{12}^2)\, e^{x_1y_2+x_2y_1}\Big\}\
\label{M4}
\ee
Other non-vanishing correlators are obtained by discrete
symmetries $1\leftrightarrow 2$, for example,
$\int U_{22}^2(U^\dagger_{22})^2\ [dU]
= \int U_{11}^2(U^\dagger_{11})^2\ [dU]$.
It is easy to check that the sum of the first and the second
lines, which should be $\int U_{11}U_{11}^\dagger [dU]$,
does indeed reproduce ${\cal M}_{11}$ from (\ref{IZduN2}).
Similarly, the sum of the second and fifth line reproduce
${\cal M}_{12}$, while that of the the second and fourth lines
should -- and does -- vanish.
Furthermore,
in the limit $x,y\longrightarrow 0$ expressions (\ref{M4})
turn into (\ref{UUU+U+}), though, as we already mentioned,
this limit is not very easy to take: even in this simple
situation ($N=2$) exponents should be
expanded up to the third order in their arguments.

Actually, this last criterium considerably restricts the
possible form of the $2m$-point correlator:
\be
\int U^{\otimes m} (U^\dagger)^{\otimes m}\ [dU]
= \frac{1}{(x_{12}y_{12})^{m+1}}
\Big( P(x_{12}y_{12})\,e^{x_1y_1+x_2y_2}
- Q(x_{12}y_{12})\,e^{x_1y_2+x_2y_1}\Big)
\ee
where $P(t) = \sum_{k=0}^m P_kt^k$ and $Q(t) = \sum_{k=0}^m Q_kt^k$
are polynomials of $t=x_{12}y_{12} = (x_1-x_2)(y_1-y_2)$ of degree $m$.
Then the condition that this expression is non-singular at $t=0$
implies that
\be
P_k = \sum_{j=0}^k \frac{(-)^j}{j!}\,Q_{k-j}
\ee
i.e. $P(t)$ is fully defined for a given $Q(t)$.
The choice of $Q(t)$ depends on the correlator. In particular,
\be
\int U_{11}^{m} (U^\dagger_{11})^{m}\ [dU] =
\frac{(-)^m}{(x_{12}y_{12})^{m+1}}
\left(\Big( \sum_{k=0}^m \frac{(-)^j m!}{j!}(x_{12}y_{12})^k\Big)
\,e^{x_1y_1+x_2y_2}
- m!\,e^{x_1y_2+x_2y_1}\right)
\ee
It is clear from this example that generalization of (\ref{Mofo})
and (\ref{Mofores}) to 4-point and higher correlators
should be straightforward, still it remains to be found.

\section*{Acknowledgements}

I am indebted to
A.Alexandrov, A.Mironov, A.Popolitov and Sh.Shakirov
for collaboration and help.

My work is partly supported by Russian Federal Nuclear Energy
Agency, by the joint grants 09-02-91005-ANF, 09-02-90493-Ukr,
09-02-93105-CNRSL and 09-01-92440-CE,
by the Russian President's Grant of
Support for the Scientific Schools NSh-3035.2008.2
and by RFBR grant 07-02-00645.


\newpage

\begin{thebibliography}{12}


\bibitem{MAMOs}
E.Wigner, Ann.Math. {\bf 62} (1955) 548; \\
F.Dyson, J.Math.Phys. {\bf 3} (1962) 140, 157,166, 1191, 1199; \\
E.Brezin, C.Itzykson, G.Parisi and J.-B.Zuber, \emph{Planar diagrams},
Comm.Math.Phys. {\bf 59} (1978) 35;\\
D.Bessis,
Comm.Math.Phys. {\bf 69} (1979) 147;\\
D.Bessis, C.Itzykson and J.-B.Zuber, Adv. Appl. Math.
{\bf 1} (1980) 109; \\
M.-L. Mehta,
Comm.Math.Phys. 79 (1981) 327;
{\it Random Matrices}, 2nd edition, Acad. Press., N.Y., 1991;\\
D.Bessis, C.Itzykson and J.-B.Zuber,
Adv.Appl.Math. {\bf 1} (1980) 109;\\
A.Migdal, \emph{Loop equations and 1/N expansion}, Phys.Rep. {\bf 102} (1983) 199; \\
F.David,
Nucl. Phys. \textbf{B257} [FS14] (1985) 45, 543; \\
J. Ambjorn, B. Durhuus and J. Frohlich,
Nucl. Phys. \textbf{B257} [FS14] (1985) 433; \\
V. A. Kazakov, I. K. Kostov and A. A. Migdal,
Phys. Lett. 157B (1985) 295; \\
D.Boulatov, V. A. Kazakov, I. K. Kostov and A. A. Migdal,
Phys. Lett. \textbf{B174} (1986) 87;
Nucl. Phys. \textbf{B275} [FS17] (1986) 641; \\
V.Kazakov,
Phys. Lett. \textbf{A 119} (1986) 140,
Mod.Phys.Lett. {\bf A4} (1989) 2125; \\
E.Brezin and V.Kazakov,
Phys. Lett. \textbf{B236} (1990) 144; \\
D.Gross and A.Migdal,
Phys. Rev. Lett. \textbf{64} (1990) 127;
Nucl.Phys. \textbf{B340} (1990) 333; \\
M.Douglas and S.Shenker, Nucl.Phys. {\bf B335} (1990) 635; \\
L.Alvarez-Gaume,
\emph{Random surfaces, statistical mechanics, and string theory},
Lausanne lectures,  1990; \\
M.Douglas, Phys.Lett. {\bf B238} (1990) 176; \\
A.Levin and A.Morozov,
Phys.Lett. \textbf{243B} (1990) 207-214; \\
P.Ginsparg, \emph{Matrix Models of 2d Gravity}, hep-th/9112013; \\
D.Gross and M.Newmann,
Phys.Lett. {\bf B266} (1991), 291; \\
A.Marshakov, A.Mironov and A.Morozov,
Phys.Lett. {\bf B 265} (1991) 99-107; \\
R.Dijkgraaf, G.Moore and R.Plesser,
Nucl.Phys. {\bf B394} (1993) 356-382,hep-th/9208031; \\
J.-M. Daul, V.A. Kazakov and I.K. Kostov,
Nucl. Phys. \textbf{B409} (1993) 311; \\
M.Staudacher,
Phys.Lett. \textbf{B305} (1993) 332, hep-th/9301038; \\
J.Ambjorn, L.Chekhov, C.F.Kristjansen and Yu.Makeenko,
Nucl.Phys. {\bf B404}(1993) 127-172,
hep-th/9302014;\\
P.Di Francesco, P. Ginsparg and J. Zinn-Justin,
Phys. Rep. {\bf 254} (1995) 1-133, hep-th/9306153; \\
A.Mironov,
Int.J.Mod.Phys. {\bf A9} (1994) 4355, hep-th/9312212;
Phys.Part.Nucl. {\bf 33} (2002) 537; \\
B.Eynard, \emph{Large Random Matrices: Eigenvalue Distribution},
hep-th/9401165; \\
M.Adler, A.Morozov, T.Shiota, P.van Moerbeke,
Nucl.Phys.Proc.Suppl. \textbf{49} (1996) 201-212, hep-th/9603066; \\
G.Akemann,
Nucl.Phys. \textbf{B482} (1996) 403-430, hep-th/9606004; \\
G.Akemann, P.H.Damgaard, U.Magnea and S.Nishigaki,
Nucl.Phys. \textbf{B487} (1997) 721-738, hep-th/9609174;
Nucl.Phys. \textbf{B519} (1998) 682-714, hep-th/9712006; \\
J.Ambjorn, M.Harris and M.Weis,
Nucl.Phys. {\bf B504} (1997) 482, hep-th/9702188; \\
A.Marshakov, M.Martellini and A.Morozov,
Phys.Lett. \textbf{B418} (1998) 294-302, hep-th/9706050; \\
T.Guhr, A.Mueller-Groeling and H.A.Weidenmueller,
Phys. Rep. \textbf{299} (1998) 189--425, cond-mat/9707301; \\
V.A.Kazakov, I.K.Kostov and N.A.Nekrasov,
Nucl. Phys. \textbf{B 557} (1999) 413, hep-th/9810035; \\
M.Adler and P.van Moerbecke, \emph{The spectrum of
coupled random matrices}, Annals of Mathematics, 1999; \\
B.Eynard, {\sl Random Matrices}, 2000,
http://www-spht.cea.fr/articles\_k2/t01/014/publi.pdf;\\
A.Gerasimov, A.Morozov and K.Selivanov,
Int.J.Mod.Phys. A16 (2001) 1531-1558,
hep-th/0005053; \\
A.Okounkov,
\emph{Generating functions for intersecting numbers
on moduli spaces of curves},
math.AG/0101201; \\
P.Ormerod and C.Mounfield,
\emph{Random Matrix Theory and the Failure of Macroeconomic Forecasts},
Physica {\bf A 280} (2000), 497-504, cond-mat/0102357; \\
P.Di Francesco,
Nucl.Phys. B648 (2003) 461-496, cond-mat/0208037;\\
P.Forrester, N.Snaith and J.Verbaarschot,
J. Phys. \textbf{A36} 2859--3645, cond-mat/0303207; \\
S.Alexandrov, V.Kazakov and D.Kutasov,
JHEP \textbf{0309} (2003) 057, hep-th/0306177; \\
N.Seiberg and D.Shih,
JHEP {\bf 0402} (2004) 021, hep-th/0312170; \\
V.Kazakov and I.Kostov,
\emph{Instantons in noncritical strings from the two matrix model},
hep-th/0403152; \\
G.Akemann, Y.V.Fyodorov and G. Vernizzi,
Nucl.Phys. \textbf{B694} (2004) 59-98, hep-th/0404063; \\
P.Di Francesco, \emph{2D Quantum Gravity, Matrix Models and Graph Combinatorics}, math-ph/0406013; \\
A.Morozov, \emph{Challenges of matrix models}, hep-th/0502010; \\
M.Stephanov, J.J.M.Verbaarschot and T.Wettig,
\emph{Random Matrices}, hep-ph/0509286;\\
Y.V.Fyodorov and B.A.Khoruzhenko,
math-ph/0602032; \\
Y.V.Fyodorov,
math-ph/0602039; \\
J.Harnad, A.Orlov,
Physica \textbf{D235} (2007) 168-206, arXiv:0704.1157

\bibitem{charmamo}
A.Migdal, JETP, {\bf 42} (1975) 413; \\
C.Itzykson and J.-B.Zuber, Comm.Math.Phys. {\bf 134} (1990) 197-208;\\
P.Di Francesco and C.Itzykson, Ann.Inst.Henri Poincare,
{\bf 59} (1993) 117;\\
V.Kazakov, M.Staudacher and T.Wynter,
Comm.Math.Phys. {\bf 179} (1996) 235; \\
I.Kostov, M.Staudacher and T.Wynter,
Commun.Math.Phys. 191 (1998) 283-298, hep-th/9703189 

\bibitem{Kazcha}
V.Kazakov, \emph{Solvable Matrix Models}, hep-th/0003064

\bibitem{UFN3}
A.Morozov,
{\it Integrability and Matrix Models},
Phys.Usp. \textbf{37 }(1994) 1-55, hep-th/9303139

\bibitem{Banff} A.Morozov,
{\it Matrix Models as Integrable Systems}, hep-th/9502091

\bibitem{DV}
F. Cachazo, K. Intriligator and C. Vafa
Nucl.Phys. B603 (2001) 3-41, hep-th/0103067; \\
F.Cachazo and C.Vafa,
\emph{N=1 and N=2 Geometry from Fluxes},
hep-th/0206017; \\
R.Dijkgraaf and C.Vafa,
Nucl.Phys. \textbf{B644} (2002) 3-20, hep-th/0206255;
Nucl.Phys. \textbf{B644} (2002) 21-39, hep-th/0207106;
hep-th/0208048; \\
L.Chekhov and A.Mironov,
Phys.Lett. \textbf{B552} (2003) 293-302, hep-th/0209085; \\
R.Dijkgraaf, S.Gukov, V.Kazakov and C.Vafa,
Phys.Rev. \textbf{D68} (2003) 045007, hep-th/0210238; \\
V.Kazakov and A.Marshakov,
J.Phys. \textbf{A36} (2003) 3107-3136,  hep-th/0211236; \\
H.Itoyama and A.Morozov,
Nucl.Phys.\textbf{B657} (2003) 53-78, hep-th/0211245;
Phys.Lett. \textbf{B555} (2003) 287-295,
hep-th/0211259;
Prog.Theor.Phys. \textbf{109} (2003) 433-463, hep-th/0212032;
Int.J.Mod.Phys. \textbf{A18} (2003) 5889-5906,
hep-th/0301136; \\
S.Naculich, H.Schnitzer and N. Wyllard,
JHEP \textbf{0301} (2003) 015, hep-th/0211254; \\
B.Feng,
Nucl.Phys. \textbf{B661} (2003) 113-138, hep-th/0212010; \\
I.Bena, S.de Haro and R.Roiban,
Nucl.Phys. \textbf{B664} (2003) 45-58,  hep-th/0212083; \\
Ch.Ann,
Phys.Lett. \textbf{B560} (2003) 116-127, hep-th/0301011; \\
L.Chekhov, A.Marshakov, A.Mironov and D.Vasiliev,
\emph{DV and WDVV}, hep-th/0301071;
hep-th/0506075; \\
A. Dymarsky and V. Pestun,
Phys.Rev. \textbf{D67} (2003) 125001,
hep-th/0301135; \\
Yu.Ookouchi and Yo.Watabiki,
Mod.Phys.Lett. \textbf{A18} (2003) 1113-1126, hep-th/0301226; \\
H.Itoyama and H.Kanno,
Phys.Lett. \textbf{B573} (2003) 227-234, hep-th/0304184;
Nucl.Phys. \textbf{B686} (2004) 155-164, hep-th/0312306; \\
M.Matone and L.Mazzucato,
JHEP \textbf{0307} (2003) 015,  hep-th/0305225; \\
R.Argurio, G.Ferretti and R.Heise,
Int.J.Mod.Phys. \textbf{A19} (2004) 2015-2078, hep-th/0311066; \\
M.Gomez-Reino,
JHEP \textbf{0406} (2004) 051, hep-th/0405242; \\
K.Fujiwara, H.Itoyama and M.Sakaguchi,
Prog.Theor.Phys. \textbf{113} (2005) 429-455, hep-th/0409060;
Nucl.Phys. \textbf{B723} (2005) 33-52, hep-th/0503113;
Prog.Theor.Phys.Suppl. \textbf{164} (2007) 125-137, hep-th/0602267; \\
Sh.Aoyama,
JHEP \textbf{0510} (2005) 032,
hep-th/0504162; \\
D.Berenstein and S.Pinansky,
hep-th/0602294

\bibitem{BreH} E.Brezin and S.Hikami,
\emph{An extension of the Harish-Chandra-Itzykson-Zuber
integral}, math-ph/0208002

\bibitem{z} A.Zabrodin,
Mod.Phys.Lett. {\bf A7} (1992) 441;
cond-mat/0210331; \\
P. Wiegmann and A. Zabrodin,
hep-th/0309253 ; \\
R. Teodorescu, E. Bettelheim, O. Agam, A. Zabrodin and P. Wiegmann,
Nucl.Phys. \textbf{B704} (2005) 407-444, hep-th/0401165

\bibitem{AMM1sf}
A.Alexandrov, A.Mironov and A.Morozov,
Int.J.Mod.Phys. {\bf A19} (2004) 4127,
Theor.Math.Phys. {\bf 142} (2005) 349, hep-th/0310113

\bibitem{HIV} T.J.Hollowood, A.Iqbal and C.Vafa,
\emph{Matrix Models, Geometric Engineering and Elliptic Genera},
JHEP \textbf{0803} (2008) 069, hep-th/0310272

\bibitem{AMMhe}
A.Alexandrov, A.Mironov and A.Morozov,
Int.J.Mod.Phys.
{\bf A21} (2006) 2481, hep-th/0412099;
Fortsch. Phys. {\bf 53} (2005) 512, hep-th/0412205;\\
A.Alexandrov, A.Mironov, A.Morozov and P.Putrov,
Int.J.Mod.Phys. {\bf A} (2009), arXiv:0811.2825

\bibitem{Eyft}
B.Eynard,
JHEP \textbf{0411} (2004) 031, hep-th/0407261; \\
B.Eynard and N.Orantin,
JHEP \textbf{0612} (2006) 026, math-ph/0504058;
math-phys/0702045; \\
L.Chekhov and B.Eynard,
JHEP \textbf{0603} (2006) 014, hep-th/0504116;
JHEP \textbf{0612} (2006) 026, math-ph/0604014; \\
B.Eynard, M.Marino and N.Orantin,
JHEP \textbf{0706} (2007) 058, hep-th/0702110; \\
N.Orantin, PhD thesis, arXiv:0709.2992;
arXiv:0803.0705

\bibitem{Ey2m} B.Eynard,
JHEP \textbf{0301} (2003) 051, hep-th/0210047;
JHEP \textbf{0311} (2003) 018, hep-th/0309036


\bibitem{EyIZ}
M.Bertola and B.Eynard,
\emph{Mixed Correlation Functions of the Two-Matrix Model},
J.Phys, {\bf A36} (2003) 7733-7750, hep-th/0303161;\\
B.Eynard, \emph{A short note about Morozov's formula},
math-ph/0406063; \\
B.Eynard and A.Prats Ferrer, Comm.Math.Phys. {\bf 264}
(2005) 115-144, hep-th/0502041; \\
A.Prats Ferrer, B.Eynard, P.Di Francesco and J.-B.Zuber,
J.Stat.Phys. \textbf{129} (2009) 885-935,
math-ph/0610049; \\
M.Bertola and A.Prats Ferrer,
\emph{Harish-Chandra integrals as nilpotent integrals},
arXiv: 0801.3452; \\
M.Bergere and B.Eynard,
\emph{Some properties of angular integrals},
arXiv: 0805.4482 

\bibitem{Klemm} N.Nekrasov and A.Okounkov,
arXiv:hep-th/0306238; \\
A.Marshakov and N.Nekrasov,
JHEP \textbf{0701} (2007) 104, hep-th/0612019; \\
B.Eynard,
\emph{All orders asymptotic expansion of large partitions},
arXiv:0804.0381;\\
A.Klemm and P.Sulkowski,
\emph{Seiberg-Witten theory and matrix models},
arXiv:0810.4944;\\
P.Sulkowski,
\emph{Matrix models for 2* theories},
arXiv:0904.3064

\bibitem{AMMmt}
A.Alexandrov, A.Mironov and A.Morozov,
{\it M-theory of matrix models},
Theor.Math.Phys. {\bf 150} (2007) 179-192, hep-th/0605171;
{\it Instantons and merons in matrix models},
Physica {\bf D 235} (2007) 126-167, hep-th/0608228; \\
N.Orantin,
{\it Symplectic invariants, Virasoro constraints and Givental decomposition },
arXiv:0808.0635

\bibitem{MShHZ}
J.Harer, D.Zagier,
Inv. Math. \textbf{85} (1986) 457-485; \\
S.K.Lando, A.K.Zvonkine,
\emph{Graphs on Surfaces and Their Applications}, Springer (2003); \\
E.Akhmedov and Sh.Shakirov,
to appear in Funkts. Anal. Prilozh., arXiv:0712.2448; \\
A.Morozov and Sh.Shakirov,
\emph{Exact 2-point function in Hermitian matrix model},
arXiv:0906.0036;
\emph{Harer-Zagier correlation functions in Gaussian matrix models},
to appear

\bibitem{MScs} A.Morozov and Sh.Shakirov,
\emph{Combinatorial Solution of Hermitian Model at Low Genera},
to appear

\bibitem{GMMMO}
A. Gerasimov, A. Marshakov, A. Mironov, A. Morozov, and A. Orlov,
Nucl. Phys. \textbf{B357} (1991) 565-618; \\
Yu.Makeenko, A.Marshakov, A.Mironov and A.Morozov,
Nucl.Phys. \textbf{B356} (1991) 574; \\
S.Kharchev, A.Marshakov, A.Mironov, A.Morozov and S.Pakuliak,
Nucl.Phys. \textbf{B404} (1993) 717-750, hep-th/9208044

\bibitem{Ko} M. Kontsevich,
Funk. Anal. Prilozh., \textbf{25:2} (1991) 50-57;
Comm.Math.Phys. {\bf 147} (1992) 1-23

\bibitem{GKM}
S.Kharchev, A.Marshakov, A.Mironov, A.Morozov and A.Zabrodin,
Phys. Lett. \textbf{B275} (1992) 311-314, hep-th/9111037;
Nucl.Phys. \textbf{B380} (1992) 181-240, hep-th/9201013;
Nucl.Phys. \textbf{B397} (1993) 339-378, hep-th/9203043\\
A.Marshakov, A.Mironov and A.Morozov,
Mod. Phys. Lett. A7 (1992) 1345-1360, hep-th/9201010;
Phys.Lett. \textbf{274B} (1992) 280-288, hep-th/9201011

\bibitem{GKMdev}
M.Adler and P. van Moerbeke, Comm.Math.Phys. {\bf 147} (1992) 25;\\
P.Di Francesco, C.Itzykson and J.-B.Zuber,
Comm.Math.Phys. {\bf 151} (1993) 193-219, hep-th/9206090; \\
S.Kharchev, A.Marshakov, A.Mironov and A.Morozov,
Mod.Phys.Lett. \textbf{A8} (1993) 1047-1062,
Theor. Math. Phys. \textbf{95} (1993) 571-582, hep-th/9208046

\bibitem{virco}
M.Fukuma, H.Kawai and R.Nakayama,
Int.J.Mod.Phys. {\bf A6} (1991) 1385; \\
R.Digkgraaf, E.Verlinde and H.Verlinde,
Nucl.Phys. {\bf B348} (1991) 565; \\
A.Mironov and A.Morozov, Phys.Lett. \textbf{B252}(1990) 47-52;\\
F.David,
Mod.Phys.Lett. {\bf A5} (1990) 1019; \\
J.Ambjorn and Yu.Makeenko, Mod.Phys.Lett. \textbf{A5} (1990) 1753; \\
H.Itoyama and Y.Matsuo, Phys.Lett. \textbf{B255} (1991) 202

\bibitem{Leq}
L.Chekhov and Yu.Makeenko,
Phys.Lett. \textbf{B278} (1992) 271-278, hep-th/9202006;
Mod.Phys.Lett. \textbf{A7} (1992) 1223-1236, hep-th/9201033; \\
J. Ambjorn, C. Kristjansen and Yu. Makeenko,
Mod.Phys.Lett. \textbf{A7} (1992) 3187-3202, hep-th/9207020

\bibitem{Umo}
E.Martinec,
\emph{On the Origin of Integrability in Matrix Models},
Commun.Math.Phys. {\bf 138} (1990) 437-450,1991; \\
V.Periwal and D.Shevitz,
Phys.Rev.Lett. {\bf 64} (1990) 1326;
Nucl.Phys. {\bf B344} (1990) 731; \\
K.Demeterfi and C.Tan, Mod.Phys.Lett. {\bf A5} (1990) 1563

\bibitem{BMS}
M.Bowick, A.Morozov and D.Shevitz,
\emph{Reduced unitary matrix models and the hierarchy of tau functions},
Nucl.Phys. \textbf{B354} (1991) 496-530

\bibitem{KazMig}
V.Kazakov and A.Migdal, \emph{Induced QCD at large N},
Nucl.Phys. \textbf{B397} (1993) 214-238,1993, hep-th/9206015

\bibitem{dA}
M.Caselle, A.D'Adda and S.Panzer,
Phys.Lett. B293 (1992) 161-167, hep-th/9207086;
B302 (1993) 80-86, hep-th/9212074

\bibitem{KazMigdev}
I.Kogan, A.Morozov, G.Semenoff and H.Weiss,
Nucl.Phys. \textbf{B395} (1993) 547-580, hep-th/9208012;
Int.J.Mod.Phys. \textbf{A8} (1993) 1411-1436, hep-th/9208054; \\
M.Dobroliubov, A.Morozov, G.Semenoff and N.Weiss,
Int.J.Mod.Phys. A9 (1994) 5033-5052, hep-th/9312145

\bibitem{Mofo}
A.Morozov,
Mod. Phys. Lett. \textbf{A7} (1992) 3503-3508, hep-th/9209074

\bibitem{Sha}
S.Shatashvili,
Comm.Math.Phys. \textbf{154} (1993) 421-432, hep-th/9209083

\bibitem{GKMKM}
S.Kharchev, A.Marshakov, A.Mironov and A.Morozov,
Int.J.Mod.Phys. \textbf{A10} (1995) 2015, hep-th/9312210

\bibitem{MMS}
A.Mironov, A.Morozov and G.Semenoff,
Int.J.Mod.Phys. \textbf{A11} (1996) 5031-5080, hep-th/9404005


\bibitem{Hur}
A.Hurwitz,
Math.Ann. {\bf 39} (1891) 1-61;
Math.Ann. {bf 55} (1902) 51-60


\bibitem{HKpf}
R.Dijkgraaf,
In: {\sl The moduli spaces of curves},
Progress in Math., 129 (1995), 149-163, Brikh\"auser; \\
R.Vakil, {\sl Enumerative geometry of curves via degeneration methods},
Harvard Ph.D. thesis (1997); \\
I.Goulden and D.Jackson,
Proc.Amer.Math.Soc. \textbf{125} (1997) 51-60,
math/9903094; \\
S.Lando and D.Zvonkine,
Funk.Anal.Appl. {\bf 33} 3 (1999) 178-188;
math.AG/0303218; \\
S.Natanzon and V.Turaev,
Topology, {\bf 38} (1999) 889-914; \\
Goulden D., Jackson D.M., Vainshtein A.,
Ann. of Comb. 4(2000), 27-46, Brikh\"auser; \\
A.Okounkov,
Math.Rev.Lett. {\bf 7} (2000) 447-453, math/0004128; \\
A.Givental,
math/0108100; \\
T.Ekedahl, S.Lando, M.Shapiro, A.Vainshtein,
Invent.Math.146(2001),297-327; \\
S.Lando,
Russ.Math.Surv., {\bf 57} (2002) 463-533; \\
A.Alexeevski and S.Natanzon,
Selecta Math., New ser. {\bf 12}:3 (2006) 307-377, math.GT/0202164;
Amer.Math.Soc.Transl. {\bf 224} (2) (2008) 1-25;
Izvestia RAN, {\bf 12}:4 (2008) 3-24; \\
S.Natanzon, Russian Math.Survey {\bf 61}:4 (2006) 185-186;
arXiv:0804.0242; \\
J.Zhou,
{\it Hodge integrals, Hurwitz numbers and symmetric groups},
math.AG/0308024; \\
A.Okounkov and R.Pandharipande,
Ann. of Math. {\bf 163} (2006) 517,
math.AG/0204305; \\
T.Graber and R.Vakil,
Compositio Math.,
{\bf 135} (2003) 25-36; \\
M.Kazarian and S.Lando,
math.AG/0410388;
math/0601760; \\
M.Kazarian,
arXiv:0809.3263; \\
S.Lando,
{\it Combinatorial Facets of Hurwitz numbers},
In: {\sl Applications of Group Theory to Combinatorics}, Koolen, Kwak and Xu, Eds.
Taylor \& Francis Group, London, 2008, 109-132; \\
V.Bouchard and M.Marino,
\emph{Hurwitz numbers, matrix models and enumerative geometry},
arXiv:0709.1458;\\
A.Mironov and A.Morozov,
JHEP \textbf{0902} (2009) 024, arXiv:0807.2843


\bibitem{MSwops}
A.Morozov and Sh.Shakirov,
JHEP, {\bf 0904} (2009) 064, arXiv: 0902.2627

\bibitem{BEMS}
G.Borot, B.Eynard, M.Mulase and B.Safnuk,
arXiv:0906.1206; \\
A.Morozov and Sh.Shakirov,
\emph{On Equivalence of two Hurwitz Matrix Models},
arXiv:0906.2573

\bibitem{MMN1}
A.Mironov, A.Morozov and S.Natanzon,
arXiv:0904.4227

\bibitem{MMN2}
A.Mironov, A.Morozov and S.Natanzon,
\emph{Integrability and ${\cal N}$-point Hurwitz Numbers},
to appear

\bibitem{AMMops}
A.Alexandrov, A.Mironov and A.Morozov,
\emph{Cut-and-Join Operators, Matrix Models} \& \emph{Characters},
to appear

\bibitem{BGW} E.Brezin and D.Gross, Phys.Lett. {\bf B97} (1980) 120;\\
D.Gross and E.Witten, Phys.Rev. {\bf D21} (1980) 446

\bibitem{AMMbgw}
A.Alexandrov, A.Mironov, A.Morozov,
arXiv:0906.3305

\bibitem{HC}
Harish-Chandra, Am.J.Math.  {\bf 79} (1957) 87; {\bf 80} (1958) 241

\bibitem{IZ}
C.Itzykson and J.Zuber, J.Math.Phys. {\bf 21} (1980) 411;\\
P.Zinn-Justin and J.-B.Zuber, J.Phys. {\bf A 36} (2003) 3173-3193,
math-ph/0209019

\bibitem{DWH} B.De Wit and G.t'Hooft, 
Phys.Lett. {\bf B69} (1977) 61

\bibitem{Givdeco}
A.Givental, \emph{Semisimple Frobenius structures at higher genus},
math.AG/0008067

\bibitem{LeSm} H.Leutwyler and A.Smilga,
Phys.Rev. {\bf D 46} (1992) 5607-5632; \\
J.Verbaarschot, hep-th/9710114; \\
R.Brower, P.Rossi and C.-I.Tan, Nucl.Phys. {\bf B190}
(1981) 699; \\
T.Akuzawa and M.Wadati, J.Phys.Soc.Jap. {\bf 67} (1998) 2151

\bibitem{Frotext}
G.Frobenius, {\it Uber gruppencharakter},
Sitzberg. Koniglich Preuss. Akad.Wiss. Berlin(1896) 985-1021;
{\it The Theory of Characters and Group Representations},
Scientific Editors of Ukraine, Kharkov, 1937;\\
D.E.Littlewood, \emph{ The theory of group characters and
matrix representations of groups}, Oxford, 1958\\
F.Murnaghan, \emph{The theory of Group Representations},
Dover, New York, 1963; \\
M.Hamermesh, \emph{Group theory and its application to physical problems},
1989\\
I.G.Macdonald, \emph{Symmetric functions and Hall polynomials},
Oxford Science Publications, 1995; \\
B. Simon, \emph{Representations of Finite and Compact Groups},
Grad.Studies in Math. \textbf{10}, AMS, Providence, 1996; \\
W.Fulton, \emph{Young tableaux: with applications to
representation theory and geometry},
London Math.Soc., 1997; \\
B. Sagan,
\emph{The Symmetric Group. Representations, Combinatorial Algorithms,
and Symmetric Functions},
2nd edition, Springer-Verlag, New York, 2001

\bibitem{Ba} A.Balantekin,
J.Math.Phys. {\bf 25} (1984) 2028;
Phys.Rev. D62 (2000) 085017, hep-th/0007161

\bibitem{DHth}
M.Semenov Tyan-Shansky, Izv.AN SSSR, Physics, {\bf 40}
(1976) 562; \\
J.J.Duistermaat and G.I.Heckman, Invent.Math. {\bf 69}
(1982) 259-268; {\bf 72} (1983) 153-158; \\
A.Alekseev, L.Faddeev and S.Shatashvili,
Geometry and Physics, {\bf 3} (1989); \\
M.Blau, E.Keski-Vakkuri and A.Niemi,
Phys.Lett. {\bf B246} (1990) 92; \\
E.Keski-Vakkuri, A.Niemi, G.Semenoff and O.Tirkkonen,
Phys.Rev. {\bf D 44} (1991) 3899; \\
A. Hietamaki, A.Morozov, A.Niemi and K.Palo,
Phys.Lett. \textbf{B263} (1991) 417-424; \\
A.Morozov, A.Niemi and K.Palo,
Phys.Lett. \textbf{B271} (1991) 365-371;
Nucl.Phys. \textbf{B377} (1992) 295-338; \\
R.Szabo, \emph{Equivariant Localization of Path Integrals},
hep-th/9608068; \\
Y.Karshon, \emph{Lecture Notes on Group Action on Manifolds},
1996-97; \\
M.Stone, Nucl.Phys. {\bf B314} (1989) 557-586

\bibitem{BeIZ}
F.Berezin and F.Karpelevich,
Doklady Acad.Nauk SSSR, {\bf 118} (1958) 9; \\
T.Guhr and T.Wettig, J.Math.Phys. {\bf 37} (1996) 6395;\\
A.Jackson, M.Sener and J.Verbaarschot, Phys.Lett.
{\bf B387} (1996) 355

\bibitem{DW} R.Dijkgraaf and E.Witten,
Comm.Math.Phys. {\bf 129} (1990) 393-429

\bibitem{UFN2}
A.Morozov, \emph{String Theory, What is it?},
Sov. Phys. Usp. \textbf{35} (1992) 671-714

\bibitem{gentau}
M. Jimbo, T. Miwa,
Publ.RIMS, Kyoto Univ,. {\bf 19} (1983) 943-1001; \\
K.Ueno, K.Takasaki,
Adv.Studies in
Pure Math., {\bf 4} (1984) 1-95; \\
A.Morozov and L.Vinet,
Int.J.Mod.Phys. A13 (1998) 1651-1708, hep-th/9409093; \\
A. Mironov, A. Morozov and L. Vinet,
Teor.Mat.Fiz. \textbf{100} (1994) 119-131,
hep-th/9312213; \\
A.Gerasimov, S.Khoroshkin, D.Lebedev, A.Mironov and A.Morozov,
Int.J.Mod.Phys. \textbf{A10} (1995) 2589-2614,
hep-th/9405011; \\
S.Kharchev, A.Mironov and A.Morozov,
q-alg/9501013; \\
A.Mironov, hep-th/9409190; Theor.Math.Phys. {\bf 114} (1998) 127,
q-alg/9711006

\bibitem{Nekin}
A.Losev, N.Nekrasov and S.Shatashvili,
Nucl.Phys. B534 (1998) 549-611, hep-th/9711108;
hep-th/9908204;
Class.Quant.Grav. 17 (2000) 1181-1187, hep-th/9911099; \\
N.Nekrasov,
Adv.Theor.Math.Phys.7:831-864,2004, hep-th/0206161;  \\
R.Flume and R.Pogossyan,
Int.J.Mod.Phys. A18 (2003) 2541, hep-th/0208176;\\
E. Frenkel, A. Losev and N. Nekrasov,
Nucl.Phys.Proc.Suppl. {\bf 171} (2007) 215-230, hep-th/0702137

\bibitem{SWTh}
N.Seiberg and E.Witten,
Nucl.Phys. {\bf B426} (1994) 19-52, hep-th/9407087; \\
A.Gorsky, I.Krichever, A.Marshakov, A.Mironov, A.Morozov,
Phys.Lett. \textbf{B355} (1995) 466, hep-th/9505035; \\
R.Donagi and E.Witten, Nucl.Phys. {\bf B460} (1996) 299, hep-th/9510101; \\
E.Martinec, Phys.Lett.{\bf B367} (1996) 91, hep-th/9510204;\\
A.Gorsky and A.Marshakov, Phys.Lett. {\bf B375} (1996) 127-134,
hep-th/9510224;\\
H.Itoyama and A.Morozov,
Nucl.Phys. B477 (1996) 855-877, hep-th/9511126;
Nucl.Phys. B491 (1997) 529-573, hep-th/9512161;
hep-th/9601168; \\
N.Nekrasov,
Nucl.Phys. B531 (1998) 323-344, hep-th/9609219; \\
A.Marshakov, A.Mironov and A.Morozov,
Phys.Lett. B389 (1996) 43-52, hep-th/9607109;
Mod.Phys.Lett. A12 (1997) 773-788, hep-th/9701014;
Int.J.Mod.Phys. A15 (2000) 1157-1206, hep-th/9701123; \\
H.Braden, A.Marshakov, A.Mironov and A.Morozov,
Phys.Lett. B448 (1999) 195-202, hep-th/9812078;
Nucl.Phys. B558 (1999) 371-390, hep-th/9902205;
hep-th/0606035; \\
A.Gorsky, A.Marshakov, A.Mironov and A.Morozov,
Phys.Lett. B380 (1996) 75-80, hep-th/9603140;
hep-th/9604078;
Nucl.Phys. \textbf{B527} (1998) 690-716,  hep-th/9802007


\bibitem{zsf}
F.Calogero, J.Math.Phys. {\bf 10} (1969) 2191, 2197; {\bf 12} (1971) 419; \\
C.F.Dunkl,
Trans.Amer.Math. {\bf 311} (1989) 167-183;
Canad.J.Math. {\bf 43} (1991) 1213-1227; \\
L.Faddeev,
Leningrad Math.J. {\bf 1} (1990) 193-225;
Alg.Anal. {\bf 1} (1989) 178-206; \\
A.Alekseev,
Nucl.Phys. {\bf B323} (1989) 719; \\
A.Gorsky and N.Nekrasov,
Nucl.Phys. B414 (1994) 213-238, hep-th/9304047;
B436 (1995) 582-608, hep-th/9401017; \\
P.Etingof, I.Frenkel and A.Kirillov,
\emph{Spherical functions on affine Lie groups},
hep-th/9407047; \\
A.Gerasimov, S.Kharchev, A.Marshakov, A.Mironov, A.Morozov
and M.Olshanetsky,
Int.J.Mod.Phys. A12 (1997) 2523-2584, hep-th/9601161; \\
P.Etingof,
\emph{Lectures on Calogero-Moser systems}, math/0606233

\bibitem{MoShID}
V.Dolotin,
\emph{On discriminants of polylinear forms}, alg-geom/9511010; \\
A.Morozov and Sh.Shakirov,
\emph{Introduction to Integral Discriminants},
arXiv:0903.2595;\\
K.Fujii,
\emph{Beyond Gaussian: A Comment},
arXiv:0905.1363


\bibitem{nolal}
I.Gelfand, M.Kapranov and A.Zelevinsky,
\emph{Discriminants, Resultants and Multidimensional Determinants},
Birkhauser, 1994; \\
V.Dolotin and A.Morozov,
\emph{Introduction to Non-Linear Algebra},
World Scientific, 2007, hep-th/0609022;
\emph{Universal Mandelbrot Set, Beginning of the Story},
World Scientific, 2006, hep-th/0501235;
Int.J.Mod.Phys.{\bf A23} (2008) 3613-3684, hep-th/0701234; \\
E.Akhmedov, V.Dolotin and A.Morozov,
JETP Lett. {\bf 81} (2005) 639-643, hep-th/0504160;\\
A. Morozov and M. Serbyn,
Theor.Math.Phys. \textbf{154} (2008) 270-293, hep-th/0703258; \\
V.Dolotin, A.Morozov and Sh.Shakirov,
arXiv:0704.2609;
Phys.Lett. \textbf{B651} (2007) 71-73, arXiv:0704.2884; \\
Andrey Morozov,
JETP Lett. {\bf 86} (2007) 745-748, arXiv:0710.2315;\\
A.Morozov and Sh.Shakirov,
arXiv:0804.4632;
arXiv:0807.4539; \\
A.Anokhina, A.Morozov and Sh.Shakirov,
\emph{Resultant as Determinant of Koszul Complex},
arXiv:0812.5013


\bibitem{BLG}
Y.Nambu,
\emph{Generalized Hamiltonian Dynamics},
Phys.Rev. {\bf D7} (1973) 2405-2414; \\
V.Filippov, Sib.Math.Jour. {\bf 26} No.6
(1985) 126; \\
L.Takhtajan,
Comm.Math.Phys. {\bf 160} (1994) 295, hep-th/9301111; \\
J.Hoppe,
Helv.Phys.Acta 70 (1997) 302-317, hep-th/9602020; \\
H.Awata, M.Li, D.Minic and T.Yoneya,
JHEP {\bf 0102} (2001) 013, hep-th/9906248; \\
J.Schwarz,
{\it Superconformal Chern-Simons Theories},
JHEP {\bf 0411} (2004) 078, hep-th/0411077; \\
A.Basu and J.A.Harvey,
Nucl.Phys. B713 (2005) 136-150, hep-th/0412310; \\
J.Bagger and N.Lambert, 
Phys.Rev. \textbf{D75} (2007) 045020, hep-th/0611108;
Phys.Rev. \textbf{D77} (2008) 065008, arXiv:0711.0955;
JHEP \textbf{0802} (2008) 105, arXiv:0712.3738;
Phys.Rev. \textbf{D79} (2009) 025002, arXiv:0807.0163; \\
A.Gustavsson, 
arXiv: 0709.1260;
arXiv: 0802.3456;\\
S.Mukhi and C.Papageorgakis,
{\it M2 to D2}, arXiv: 0803.3218; \\
M.Bandres, A.Lipstein and J.Schwarz,
{\it N=8 Superconformal Chern-Simons Theories},
arXiv: 0803.3242; \\
A.Morozov,
JHEP \textbf{0805} (2008) 076, arXiv:0804.0913;
JETP Lett. \textbf{87} (2008) 659-662, arXiv:0805.1703;\\
J.Gomis, G.Milanesi, and J.G.Russo,
arXiv:0805.1012 v2; \\
S.Benvenuti, D.Rodriguez-Gomez, E.Tonni and H.Verlinde,
arXiv:0805.1087;\\
P.-M.Ho, Y.Imamura and Y.Matsuo,
{\it M2 to D2 revisited},
arXiv:0805.1202; \\
O.Aharony, O.Bergman, D.L.Jafferis and J.Maldacena,
JHEP \textbf{0810} (2008) 091, arXiv:0806.1218; \\
I.A.Bandos and P.K.Townsend,
Class.Quant.Grav. \textbf{25} (2008) 245003, arXiv:0806.4777;
JHEP \textbf{0902} (2009) 013, arXiv:0808.1583; \\
J.Gomis, D.Rodriguez-Gomez, M.Van Raamsdonk, H.Verlinde,
JHEP \textbf{0809} (2008) 113, arXiv: 0807.1074; \\
J.A.Minahan, W.Schulgin and K.Zarembo,
arXiv:0901.1142; \\
J.Figueroa-O'Farrill,
\emph{Deformations of 3-algebras}, arXiv:0903.4871; \\
D.Kamani,
\emph{Evidence for the $p+1$-algebra for super-$p$-brane},
arXiv:0904.2721


\end{thebibliography}
\end{document}